\documentclass{emulateapj}

\newcommand{\pr}{^\prime}
\newcommand{\e}{\epsilon}
\newcommand{\g}{\gamma}
\newcommand{\gp}{\gamma}
\newcommand{\gptwo}{\gamma^2}

\newcommand{\Dop}{\delta_{\rm D}}

\newcommand{\Op}{\Omega^\prime}
\newcommand{\mup}{\mu^\prime}

\newcommand{\ep}{\epsilon^\prime}

\newcommand{\psim}{\lower.5ex\hbox{$\; \buildrel \propto \over\sim \;$}}
\newcommand{\lbar}{\lower.0ex\hbox{$\; \buildrel {\lower0.0ex \hbox{-}} \over\lambda  \;$}}
\newcommand{\es}{\epsilon_{*}}
\newcommand{\Os}{\Omega_{*}}
\newcommand{\lC}{\lambda\!\!\! \raisebox{0.5ex}{--}_{\rm C}}

\shorttitle{Gamma Rays from Microquasars}
\shortauthors{Dermer \& B\"ottcher}

\begin{document}
\title {Gamma Rays from Compton Scattering in the Jets of Microquasars: \\
Application to LS 5039}

\author{Charles D.\ Dermer\altaffilmark{1} \& Markus B\"ottcher\altaffilmark{2}}

\altaffiltext{1}{E.\  O.\  Hulburt Center for Space Research, Code 7653,
Naval Research Laboratory, Washington, DC 20375-5352}
\altaffiltext{2}{Astrophysical Institute, Department of Physics 
and Astronomy, Ohio University, Athens, OH 45701, USA}

\begin{abstract}
Recent High Energy Stereoscopic System (HESS) observations show that 
microquasars in high-mass systems are sources of very high energy
$\gamma$-rays.  A leptonic jet model for microquasar $\gamma$-ray emission
is developed.  Using the head-on approximation for the Compton cross
section and taking into account angular effects from the star's
orbital motion, we derive expressions to calculate the spectrum of
$\g$ rays when nonthermal jet electrons Compton-scatter photons of the
stellar radiation field. The spectrum of Compton-scattered
accretion-disk radiation is also derived by approximating the
accretion disk as a point source of radiation located behind the
jet. Numerical results are compared with simpler expressions obtained
using $\delta$-function approximations for the cross sections, from which
beaming factors are derived.  Calculations are presented for power-law
distributions of nonthermal electrons that are assumed to be
isotropically distributed in the comoving jet frame, and applied to
$\gamma$-ray observations of LS 5039. We conclude that (1) the TeV
emission measured with HESS cannot result only from Compton-scattered
stellar radiation (CSSR), but could be synchrotron self-Compton (SSC)
emission or a combination of CSSR and SSC; (2) fitting both the HESS
data and the EGRET data claimed to be associated with LS 5039 requires
a very improbable leptonic model with a very hard electron energy
distribution.  Because the $\gamma$ rays would be variable in a
leptonic jet model, the data sets are unlikely to be representative of
a simultaneously measured $\gamma$-ray spectrum. We therefore
attribute EGRET $\gamma$ rays primarily to CSSR emission, and HESS
$\gamma$ rays to SSC emission.  Detection of periodic modulation of
the TeV emission from LS 5039 would favor a leptonic SSC or cascade
hadron origin of the emission in the inner jet, whereas stochastic
variability alone would support a more extended leptonic model.  The
puzzle of the EGRET $\gamma$ rays from LS 5039 will be quickly solved
with GLAST.

\end{abstract}


\section{Introduction}

X-ray binaries with jets, or microquasars, are common in
our Galaxy, with $\approx 16$ now known 
\citep[for a recent review see, e.g.,][]{par05}. About
one-third are high-mass X-ray binaries (HMXBs), including Cygnus X-1,
Cygnus X-3, LS 5039, and LSI+61$^\circ$303, and the remainder are
low-mass X-ray binaries (LMXBs), including GRS 1915+105, GRO J1655-40,
Sco X-1, and 1E 1740.7-2942.  The compact companions are a mixture of
black holes and neutron stars, and the radio activity of the
microquasars is about equally divided into persistent and transient
behaviors.

Recent observations \citep{aha05} made with the High Energy
Stereoscopic System (HESS) show that the high-mass microquasar LS 5039
is a source of very high energy (VHE) $\gamma$ rays in the $\approx
200$ GeV -- 10 TeV range, confirming its earlier tentative
identification with the EGRET source 3EG J1824-1514 \citep{par00}. 
A second high-mass microquasar system, LSI+61$^\circ$303 (V615 Cas), is
associated with the COS-B source 2CG 135+01 \citep{her77,gt78} and the
EGRET source 3EG J0241+6103 \citep{kni97}, but is too far north for
observations with HESS. 
The EGRET source 3EG J1824-1514 associated with LS 5039 shows marginal
evidence for variability \citep{tor03}.  In contrast, the EGRET light
curve of 3EG J0241+6103, the counterpart to LSI+61$^\circ$303, is
strongly variable \citep{tav98}.  Moreover, \citet{mas04} performed a
timing analysis of the 3EG J0241+6103 data and found a most probable
period of $27.4\pm 7.2$ days, compared to its 26.5 day orbital period.

Evidence for stochastic and periodic variability of these sources at
$\gamma$-ray energies would argue in favor of a leptonic microquasar
jet model similar to blazar jet models \citep[for recent reviews of
microquasar models, see][]{rom05,fm04}, especially if the X-ray and
$\gamma$-ray emissions display correlated variability \citep{gbd05}. 
The importance of Compton scattering of external photons to produce
gamma-rays in blazar jets was first
considered by \citet{bs87} and \citet{mk89} and later, in view of 
the {\it Compton Observatory} discoveries, by \citet{dsm92} and \citet{sbr94}.
A microquasar jet model differs importantly from a blazar jet model
through the addition of the stellar radiation field from the high-mass
star and the periodic orbital modulation of the binary stellar system
\citep{gak02,krm02}. Although
\citet{cas05} claim low significance periodic variability when folding
the HESS data for LS 5039 with its orbital period, the EGRET 
data showed no compelling evidence for either stochastic or periodic
variability \citep{par00}. The X-rays from LS 5039 are, however, 
moderately variable. {\it RXTE} observations in the 3 -- 30 keV
range may show periodic variability correlated with the
periastron passage of LS 5039, so it is unceratin whether 
the X-rays are associated with
accretion disk or the jet \citep{bos05}.

In addition to stellar and accretion-disk emissions, microquasar
emission from the jet will produce a variable multiwavelength
continuum consisting of radio/IR and jet X-ray \citep{mff01}
synchrotron radiation.  Nonthermal $\gamma$-ray emission is likely to
originate from synchrotron self-Compton (SSC) \citep{aa99} and
external Compton (EC) processes \citep{lb96,gak02} by these same jet
electrons.  The bright high-mass star makes an important contribution
to the external radiation field in HMXBs, whereas the accretion disk
is the dominant external photon source in LMXBs
\citep{gre05}. Compton-scattering leptonic jet models 
of the $\gamma$-ray emission
from LS 5039 and LSI+61$^\circ$303 are presented by
\citet{bp04a,bp04b}. 

In this paper, we perform a Compton-scattering analysis of the jet
$\gamma$-ray emission from HMXB microquasars for a leptonic jet model,
focusing on Compton-scattering effects from the azimuthal variations
of the stellar radiation field using parameters inferred from observations
\citep{mcs01,cas05} of LS 5039, which has a period
of $3.90603\pm 0.00017$ days.  We assume that the twin jets of the
microquasar are oriented normal to the orbital plane of the compact
object and star; geometrical complications of precessing jets are not
considered here.  The orbital variations of the bright O or B stars
introduce interesting kinematic variations that appear in the
$\gamma$-ray emission spectrum if the $\gamma$ rays are due to stellar
photons that are Compton-scattered by nonthermal jet electrons,
including variations of peak $\nu F_\nu$ photon energy and inferences
of the locations of the $\gamma$-ray emission site.  This emission is 
also subject to the effects of $\gamma\gamma$ absorption \citep{bd05,dub05},
although this effect is not included in the calculations shown here.

Angle-dependent effects on Compton-scattered jet radiation are treated
in Section 2.  Approximations made in the derivation are clearly
enumerated, so that they can be relaxed in more detailed numerical
treatments. In particular, the analysis employs a fixed electron
distribution. Spectral calculations are presented in Section 3 using
parameters appropriate to LS 5039. Difficulties to fit the combined
EGRET and HESS spectra of LS 5039, if assumed to be simultaneously
radiated, are discussed in Section 4. 
Implications for establishing the nature of microquasar
$\gamma$-ray emission from LS 5039 from further HESS and upcoming
GLAST observations are also considered. A summary of the results 
is given in Section 5.

 \section{Gamma Rays from Compton-Scattered Stellar Radiation}

The geometry of the microquasar system is shown in Fig.\ 1.  The
generic system considered here is a HMXB with parameters taken from
observations of LS 5039, but our results are also applicable to LMXBs
when the accretion disk is approximated by a hot spot at the base of
the twin jets.  The star and the compact object, separated by distance
$d$, are assumed to follow circular orbits around their common center
of mass.  Material from Roche-lobe overflow in low-mass systems, or
stellar winds in high-mass systems, forms an accretion disk
surrounding the compact object. As the matter accretes onto the
neutron star or black hole, plasma is assumed to be ejected transverse
to the orbital plane in the form of twin jets.

\begin{figure}[t]
\vskip-0.5in
\epsscale{1.1}
\plottwo{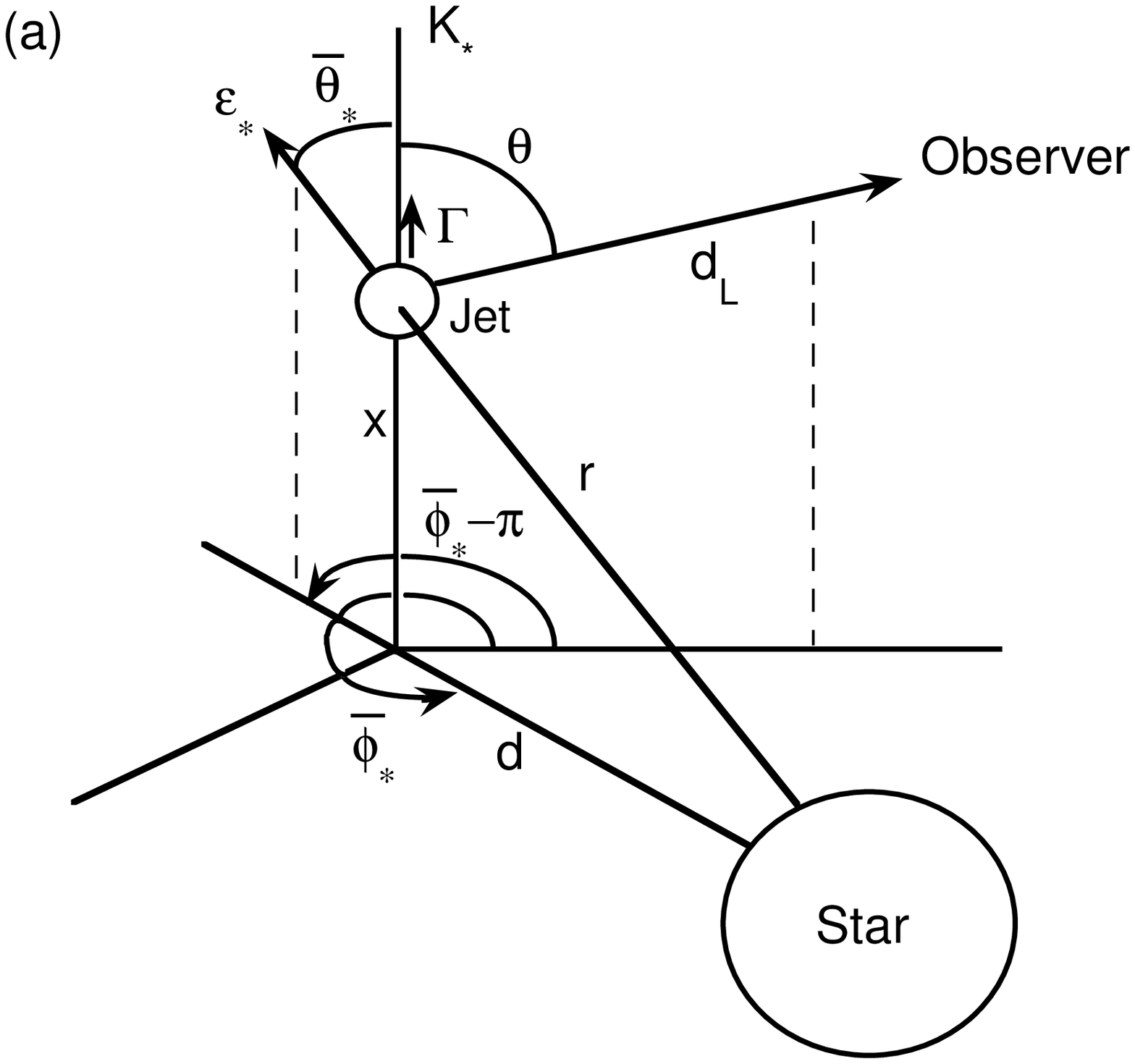}{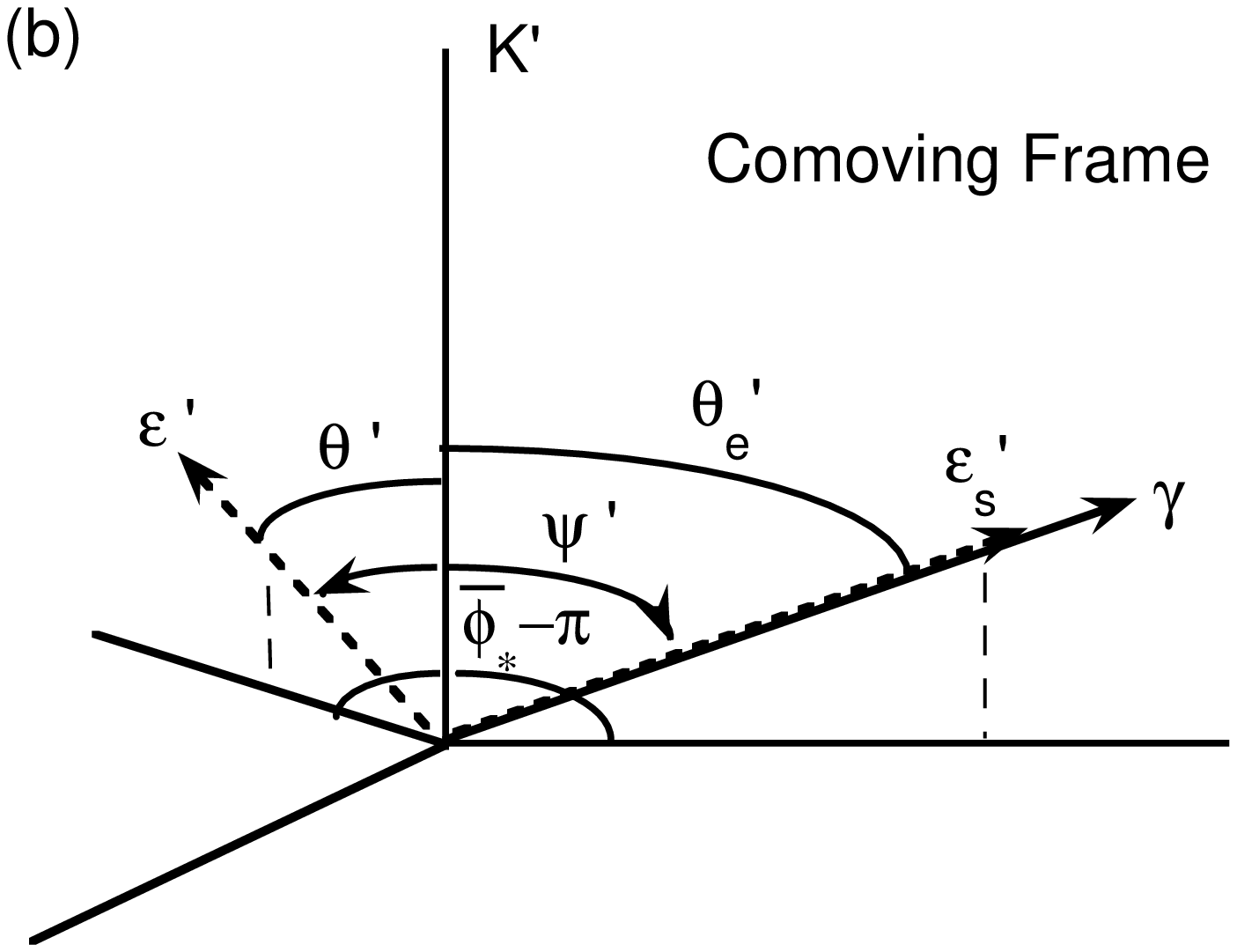}
\caption{Geometry of scattering in the microquasar system.
 (a) Stationary frame $K_*$ of the microquasar system. The stellar
 azimuthal angle $\bar\phi_* = 0$ when the high-mass star is closest
 to the observer. The photon azimuth in the comoving frame is then
 $\bar\phi_* - \pi$.  (b) Comoving frame $K^\prime$ of the jet
 plasma. In the head-on approximation, the photon is
scattered along the direction of the scattering electron. See 
text for definition of symbols.}
\end{figure}

\subsection{Model Assumptions}

Given that the dominant radiation mechanism for microquasar
$\gamma$-ray emission is undecided, for example, whether microquasar
emissions have a primary hadronic \citep[e.g.,][]{romero03,aha05a} or
leptonic \citep[e.g.,][]{brp05,pbr05} origin, an overly complicated
leptonic jet model for microquasar emission seems unjustified at this
time.  A number of simplifying assumptions are therefore made that can
be relaxed in more detailed treatments. Specifically, we assume that
\begin{enumerate}
\item the jets are steady and oriented in the direction normal 
to the orbital plane 
of the binary system, with the jet electrons confined to the 
axis defining the jet direction;
\item the orbit is circular;
\item the star and accretion disk can be treated as point sources of radiation;
\item the nonthermal electrons are isotropically distributed in the comoving
frame with a fixed energy distribution, and most of the radiation is
emitted at a specified distance above the orbital plane; and
\item cascade processes can be neglected.
\end{enumerate}

Assumption 1 means that we neglect additional periodicities of the jet
ejection process possibly associated with periastron passage, or disk
precession that could lead to additional modulation of the
$\gamma$-ray signal. This latter periodicity
\citep[e.g.,][]{krm02,tor05} could be different 
from the orbital periodicity. Moreover, the jets need not be aligned
normal to the orbital plane.  For example, the jets in V4641 Sgr may
lie within $\sim 36^\circ$ of the orbital plane \citep{but03}. These
effects can be dealt with in the present model by treating a more
complicated geometry, and X-ray and TeV observations should be
cross-correlated to search for periods unrelated to orbital motions.
Helical motions of jet trajectories, which
may operate in black-hole jet systems \citep[e.g.,][]{rei04}, would
also require a more detailed treatment than considered here.

Regarding assumption 2, scattering kinematics for highly eccentric orbits
can be developed on the basis of the treatment presented here.  LS
5039 executes a moderately elliptical orbit in a system with a 22.9 $M_\odot$ 
star of spectral type O6.5, with eccentricity $\cong 0.35$ \citep{cas05}.  
This effect will be displayed in both 
the scattering kinematics \citep{pbr05} and $\gamma\gamma$ absorption
calculations \citep{dub05}, but does not, to first order, change the
conclusions we draw by studying anisotropic Compton scattering and
$\gamma\gamma$ absorption for circular orbits.

For the purposes of modeling VHE $\gamma$-rays, analysis of a
microquasar system must take into account the angle-dependence of the
stellar and accretion-disk radiation field both in the Thomson and
Klein-Nishina (KN) regimes of scattering. Although \citet{gkm01}
provide a useful method to calculate scattered radiation spectra in
the KN regime, it is applicable as presented only to surrounding
isotropic radiation fields, and for jets with fixed isotropic electron
distributions.
\citet{pbr05} and \citet{brp05} also treat the anisotropy of 
the stellar radiation field
in a simplified fashion, without fully taking into account the effects 
of the directional target photon field.

In this work, we treat the angular dependence of Compton-scattered
stellar and accretion-disk radiation fields employing an accurate
approximation to the Compton cross section in the head-on
approximation that is valid throughout the Thomson and KN
regimes.  To avoid over-complicated expressions, the star and
accretion-disk are treated as point sources of photons scattered by
nonthermal electrons in a relativistically-moving jet with negligible
extent (assumption 3).  These assumptions simplify the derivation, but
in no way violate the essential geometry of the microquasar system,
and are straightforward (if tedious) to relax.  Nevertheless, it is
important to note that the companion star in a HMXB can subtend as
much as $\approx 10$\% of the full sky as seen from a location close
to the compact object.

We also assume a non-evolving electron distribution and calculate
emission from a fixed jet location (assumption 4). As the jet plasma
moves away from the microquasar system, the electrons are energized,
for example, through internal collisions of ejected plasma shells or
via shocks formed during interactions of the jetted plasma with the
external medium. The nonthermal electrons subsequently lose energy
through expansion and radiative losses. In more detailed treatments,
electron-energy evolution through synchrotron, Compton, and adiabatic
losses are considered \cite[e.g.,][]{aa99,bos05a,gbd05}. When stellar
radiation fields are important, a consistent treatment of the
evolution of the nonthermal electron Lorentz factor distribution with
location requires, in addition to the synchrotron, SSC, and adiabatic
energy-loss rates, electron energy-loss rates obtained from the
Compton-scattered stellar and accretion-disk radiation fields, as
derived here. Given that $\gamma$-ray telescopes integrate over a long
time to accumulate signal in comparison with Compton and synchrotron
cooling time scales for the $\gamma$-ray emitting electrons, as is
easily demonstrated, the detected emission is well-approximated as the
radiation from a time-averaged electron distribution. Yet the
power-law form of the electron distribution remains an assumption that
may not agree with calculations of electron evolution.

Finally, we do not treat the electromagnetic cascades formed through
$\gamma\gamma$ attenuation as $\gamma$-rays propagate through the
anisotropic radiation field of the jet and star \citep[see,
e.g.,][]{pmd92,bed97,aha05a}. Cascading effects could be important
when the absorption depth to $\gamma\gamma$ pair production
attenuation is large, namely within the inner jet\footnote{Inner jet
refers to jet locations $x \lesssim d$.} \citep{bd05,dub05}.  This
effect could, in principle, reduce the level of modulation of the VHE
$\gamma$-ray signal from $\gamma\gamma$ attenuation and introduce new
spectral components.  Relaxing assumption (5) introduces, however, new
assumptions about the strength and geometry of the magnetic field in
the vicinity of the microquasar outside the jet plasma that are
difficult to constrain.

\subsection{Stellar Radiation Field}

We treat a system (Fig.\ 1a) where the jet outflow has constant bulk
Lorentz factor $\Gamma = 1/\sqrt{1-\beta^2}$.  In the stationary frame
of the microquasar system (where the center-of-mass of the compact
object and high-mass star is at rest), the plasma will have traveled a
distance $x=\beta c t_{*}$ from the compact object and reached a
distance $r = \sqrt{x^2+d^2}$ from the companion star after time
$t_*$, measured from the moment of ejection of the jetted plasma at
$x= 0$.

Let $\bar\phi_{*}(t) = \omega_{*}t_{*}$ represent the star's orbital
phase measured from the angle when the high-mass star is closest to
the observer, where $\omega_{*} = 2\pi/P$ is the star's angular
frequency and $P$ is the period (from hours to days for LMXBs, and
days to weeks for HMXBs).  Photons from the companion star impinge on
the outflowing jet plasma at an angle $\bar\theta_* = \arccos \bar
\mu_*$ to the jet axis as measured in the stationary frame of the
star, where $\bar\mu_*= x/r = 1/ \sqrt{1+d^2/ x^2}$.  The angle
$\theta = \arccos \mu$ is the inclination of the observer with respect
to the jet axis.

The energy flux at distance $r$ from a uniform brightness sphere
(i.e., the star) of radius $R_*$ is $${d{\cal E_*}\over dA dt d\es} =
{\pi R_*^2\over r^2} B_{\e_*}\;,$$ where ${\cal E}_*$ is the radiated
photon energy and $\es = h\nu_* /m_ec^2$ is the dimensionless photon
energy. The intensity $$B_{\e_*} = I_{\e_*}^{bb}(\Theta ) = {2 m_e c^3
\es^3\over \lambda_{\rm C}^3 [\exp(\es/\Theta) - 1]}$$ for a
blackbody, where $\lambda_{\rm C}=h/m_e c= 2.42\times 10^{-10}$ cm is
the Compton wavelength of the electron and $\Theta = k_{B}T_*/m_ec^2$
is the dimensionless temperature of the star.

Consider a star with luminosity $L_*$ and temperature $T_*$, so that
the stellar radius $R_* = \sqrt{L_*/4\pi \sigma_{\rm SB}T^4_*}$, and
$\sigma_{\rm SB}$ is the Stefan-Boltzmann constant. The differential
energy density $$u^*_{bb}(\e_{*};r)= {1\over c} {d{\cal E}_*\over dA
dt d\e_*}.$$ Hence
\begin{equation}
u^*_{bb}(\e_{*},\Omega_{*};r)= u^0_*\;{\e_*^3 
\delta(\mu_*- \bar\mu_*) \delta(\phi_* - \bar \phi_*)
\over \exp(\es/\Theta) - 1 }\;,
\label{ustarbb}
\end{equation}
where 
$$u^0_* = {15L_*\over 4\pi^5 c\Theta^4 r^2}\;.$$
The stationary-frame photon density
$n^*_{ph}(\es,\Os)=u^*(\es,\Os)/(m_ec^2\es)$. 

The $f^*_\e = \nu F^*_\nu$ spectrum of the star measured by an
observer located a (luminosity) distance $d_L$ away from the star is
\begin{equation}
f^*_\e = {15L_*\over4\pi^5 \Theta^4d_L^2}\; 
{\e^4\over \exp(\e/\Theta ) - 1}\;,
\label{festar}
\end{equation}
where $\e = \e_*/(1+z) \cong \e_*$ for the low redshift ($z \ll 1$)
sources considered here.

\subsection{Stellar Radiation Photons Compton-Scattered by Jet Electrons} 

The detailed derivation of the Compton-scattered stellar radiation (CSSR) spectrum
is given in Appendix A. The $\nu F_\nu$ spectrum resulting from Compton-scattered stellar
radiation (CSSR) for a uniform emitting region filled with 
an isotropic comoving distribution of electrons is
$$f_\e^{\rm C*} = {3 c\sigma_{\rm T} u_*^0  \Dop^2\e^2\over 32\pi d_L^2}\;
\int_{\e/\Dop}^\infty d\gp\;
{N^\prime_e(\gp )\over \gptwo}\times$$
\begin{equation}
a^2\big[g^{-2}\;(y + {1\over y})I_1 - {2\e\over \Dop \gp y b g}I_2
+ \big({\e\over \Dop \gp y b}\big)^2 I_3\;\big]\;
\label{feC_1}
\end{equation}
where $y = 1-(\e/\Dop\g)$,  
\begin{equation}
\Dop = [\Gamma(1-\beta \mu)]^{-1}\;,
\label{Doppler}
\end{equation}
 is the Doppler factor,
$\e = h\nu/m_ec^2$ is the dimensionless observed photon energy, $N^\prime_e(\gamma)$ is differential
distribution of electrons with comoving Lorentz factors $\gamma$,
and $d_L$ is the luminosity
distance to the source.  The functions a, b, $I_1$, $I_2$ and $I_3$ are 
defined in Appendix A.

\subsection{Compton Spectrum from Point Source of Radiation Field Behind Jet}

The accretion disk provides a source of external radiation that enters
the jet from behind. For a Shakura-Sunyaev accretion disk, the
distance where the transverse extent of the accretion disk can be
neglected and the accretion disk can be approximated by a point source
is given by
\begin{equation}
 x \gg \Gamma^4 r_g\;
\label{xG4}
\end{equation}
\citep{ds93,ds02}, where $r_g = GM/c^2$ is the gravitational 
radius of the black hole with mass $M$.
For the mildly relativistic speeds ($\Gamma \approx 1$ -- 2)
considered here, and jet distances of order of the orbital radius, the
accretion disk can be well approximated as a point source of
radiation. The accretion-disk radiation spectrum may not be well
represented by a blackbody or a Shakura-Sunyaev spectrum, as
accretion disks display a wide range of spectra. This does not affect
eq.\ (\ref{xG4}), as this estimate is based on energy dissipation
at various radii, which is largely unaffected by disk type.

Following the procedure in Appendix A, again using the
head-on approximation, eq.\ (\ref{dsigC}), for the Compton cross section, but
now for a point source at the origin, we derive the $\nu F_\nu$ spectrum of
Compton-scattered accretion disk (CSAD) radiation given by
$$f_\e^{pt} = {3\e^2 \Dop^2\sigma_{\rm T} \over 
128\pi^2 x^2 d_L^2 }\int_0^\infty d\e_0 \; {L_0(\e_0 )\over \e_0^2}
\;\int_{\gp_{min}}^\infty d\gp\;
{N^\prime_e(\gp )\over \gptwo}\times$$
\begin{equation}
\;\big[ y + y^{-1} - 
{2\e\over \Dop \gp \hat\e_i y} +
\big({\e\over \Dop\gp\hat\e_i y}\big)^2\big]\;,
\label{fept}
\end{equation}
where $\hat \e_i = \gamma \e_0 \Dop (1-\mu)$ and $$\gp_{min} = {\e\over 2\Dop}
\;\big[ 1 + \sqrt{1 + {2\over \e\e_0(1-\mu)}} \;\big]\;.$$ Appendix B
gives approximate expressions for the CSSR and CSAD spectra in the
Thomson and KN regimes that reduce the number of numerical
integrations, though at the expense of accuracy.  These simpler
expressions have, however, the virtue of allowing the beaming factors
of the various processes to be simply derived.

\section{Results}

We calculate the CSSR spectrum using the standard parameters listed in
Table 1.  The standard jet height $x$ is taken to be equal to the mean
orbital separation $d = 2.5\times 10^{12}$ cm of LS 5039. We also
employ a broken power-law distribution for the nonthermal electrons,
given by
\begin{equation}
N_e^\prime (\g ) = K [\g_1^{q-p}\g^{-q}H(\g;\g_0,\g_1) + 
\g^{-p}H(\g;\g_1,\g_2 )]\;.
\label{elecbpl}
\end{equation}
Normalizing to the total comoving nonthermal electron energy
$$W_e^\prime \;\cong\; m_ec^2 \int_1^\infty d\g\; \g\;N_e^\prime (\g
)$$ gives
\begin{equation}
 K = {W_e^\prime\over m_e c^2} \;
\big[{\g_1^{q-p}(\g_1^{2-q} - \g_0^{2-q})\over 2-q} 
+ {\g_1^{2-p}-\g_2^{2-p}\over p-2}\big]^{-1}\;.
\label{K}
\end{equation}
The maximum electron Lorentz factor is calculated from the well-known
\citep{gfr83} radiation-reaction limit obtained by equating
the electron energy-loss timescale with the gyration timescale $t_g =
m_e c\g/eB$, which holds provided that synchrotron losses dominate
Compton-energy losses. The situation is more complicated here, where
Compton-losses from the stellar radiation field are large and 
KN effects are important \citep{aha05a}.
In Appendix C, an improved treatment of this limit for relativistic
jets is performed, and we quantify the regime where the synchrotron
radiation-reaction limit holds. We take $\g_2 = \g_{max}$ here, with
$\g_{max}$ given by eq.\ (\ref{gammamaxsyn}) and $\eta = 1$, giving
the most optimistic maximum synchrotron frequency.

\begin{figure}[t]
\vskip-0.5in
\epsscale{1.1}
\plottwo{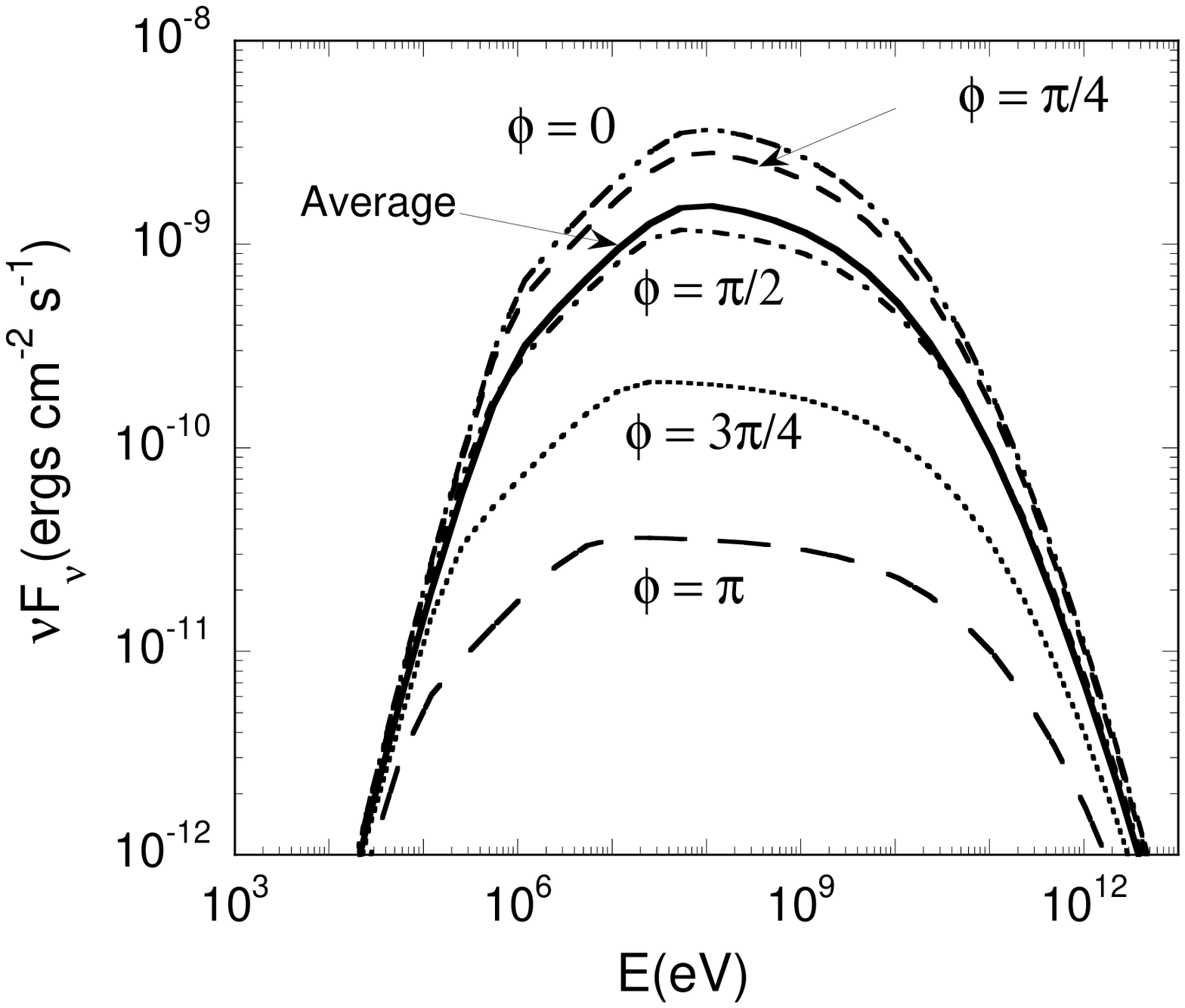}{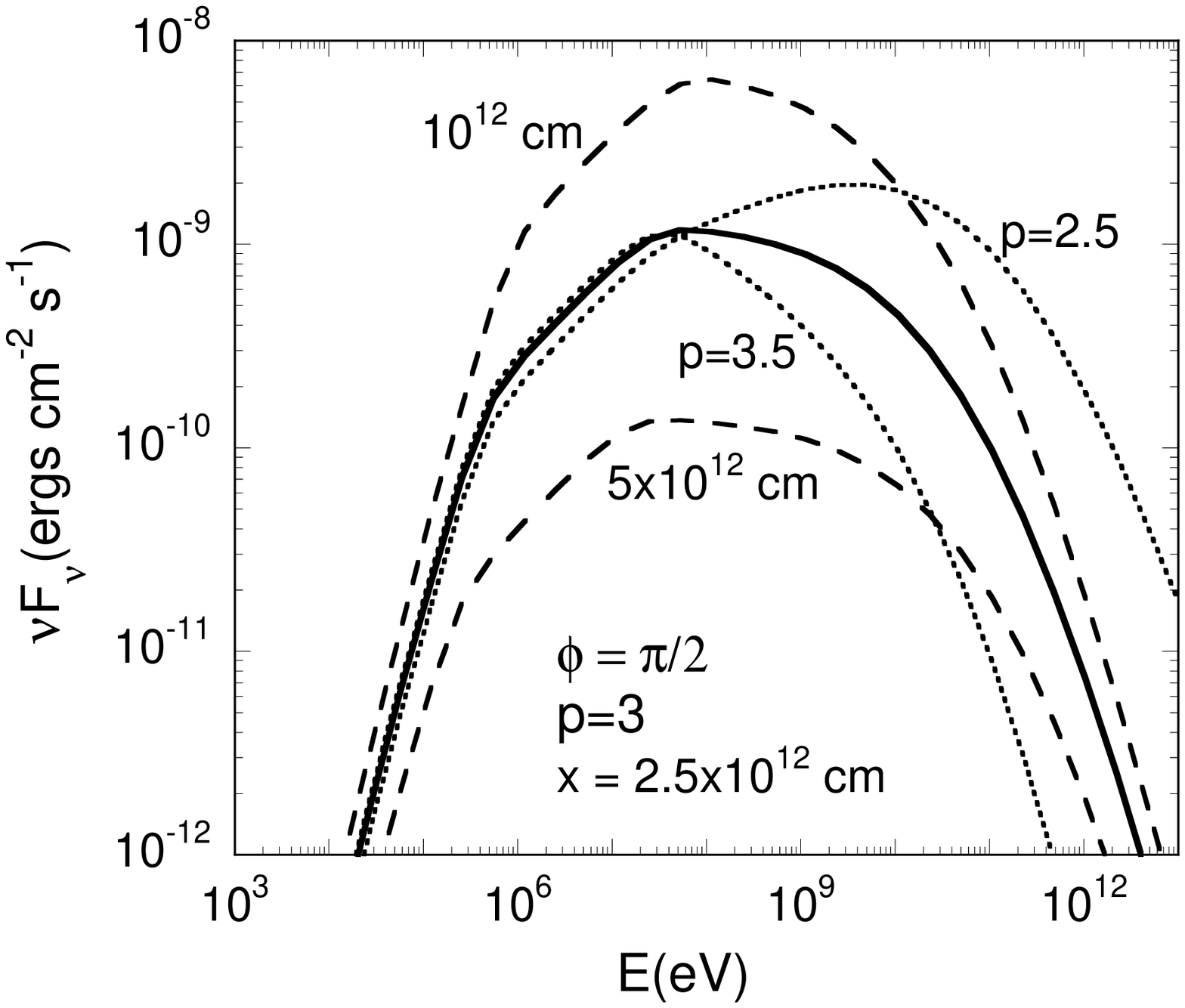}
\caption{The $\nu F_\nu$ Compton-scattered 
stellar radiation (CSSR) spectral energy distribution (SED) from a
microquasar jet when nonthermal jet electrons Compton-scatter photons
of the high-mass companion star.  Standard parameters of the system,
appropriate to LS 5039, are given in Table 1.  In particular, the jet
height above the orbital plane of the system is assigned a standard
value of $x = 2.5\times 10^{12}$ cm, the jet plasma outflow Lorentz
factor $\Gamma = 2$, and the number index of the electron energy
distribution is $p = 3$.  (a) Dependence of the SED on phase $\phi$,
with $\phi = 0$ when the companion star is nearest the observer (Fig.\
1a). The average SED integrated over all phases is shown by the solid
curve. (b) Dependence of the CSSR SED on $x$ and $p$ with $\phi =
\pi/2$. The solid curve shows the SED for the standard parameters, and
the other curves are labeled with the parameter that has changed from
the standard parameter values. Effects of $\gamma\gamma$ attenuation
on the SEDs are not included here.}
\end{figure}

Fig.\ 2 show calculations of the $\nu F_\nu$ SED of the CSSR process.
The dependence of the SED on phase of the binary orbit is shown in
Fig.\ 2a, and the dependence of the SED on various parameters of the
binary system at $\phi = \pi/2$ is shown in Fig.\ 2b (we now simplify
the notation by letting the stellar azimuth angle $\bar \phi_* \rightarrow
\phi$). In both figures,
the steep lower energy behavior is a consequence of the low-energy
cutoff Lorentz factor $\g_0 = 100$ assumed in the electron
distribution. The energy of this low-energy cutoff, and the cutoff
associated with the change in slope at $\g_1 = 10^3$, occurs for the
blackbody stellar spectrum around photon energy
\begin{equation}
E_{\rm T} \cong 2.7 k_{\rm B} T_* \g_i^2 \Dop
\Gamma(1-\beta\bar\mu_*) (1-\cos\bar\psi^\prime)\;,\;i = 0, 1\;,
\label{EgbreakT}
\end{equation}
provided that the scattering takes place in the Thomson regime.  (This
result is found from the $\delta$-function approximations for Thomson
scatering given in Appendix B.) Except for the phase-dependent
factor $(1-\cos\bar\psi^\prime)$, all the terms in eq.\
(\ref{EgbreakT}) are uniquely determined. For the standard parameters
of Table 1, $\beta = 0.866$, $\Dop = 2.32$ and $\bar\mu_* = 0.707$,
so that the breaks due to Thomson scattering occur at $E_\g \cong 16
\g_i^2 (1-\cos\bar\psi^\prime)$ eV, in accord with the results
shown. The breaks in the photon spectrum in Fig.\ 2a occur at lower
energies when $\phi= \pi$ than when $\phi = 0$ because the $\phi =
\pi$ case involves more nearly tail-on scattering events that scatter
the target photons to lower energies than for the more nearly head-on
scattering events with $\phi \approx 0$ (see Fig.\ 1a).

The spectral index of the Compton-scattered stellar radiation follows
the well-known behavior $$\alpha_\nu = {3-p\over 2}\;$$ in the Thomson
regime, as shown in Appendix B, where $\alpha_\nu$ is the $\nu
F_\nu$ spectral index\footnote{At lower energies, $\alpha_\nu
\cong 2$, corresponding to the low-energy emissivity spectrum 
$j^\prime \propto \ep$ of a mono-energetic electron  distribution 
\citep{bg70}.}.
For electron index $p = 2$ assumed in the lower branch of the electron
spectrum in Fig.\ 2a, the spectrum rises with slope $1/2$ in a $\nu
F_\nu$ plot.  For the upper branch of the electron spectrum with $p =
3$, the spectrum is flat, as seen for the $\phi = \pi$ curve in Fig.\
2a. For the more nearly head-on scattering events with $\phi \approx
0$, the transition to the KN regime occurs at lower photon energies,
so that scattering in the Thomson regime is never fully
achieved. Indeed, the CSSR spectral index at high-energies
asymptotically approaches the steep KN index $\alpha_\nu \approx 1-p$
(eq.\ [\ref{A7}]).

Transition to the KN regime of scattering, as derived in Appendix B, 
takes place over a broad range of photon
energies centered around\footnote{The factor 1/4 is introduced because
the domain of Compton scattering is defined by the value of the
quantity $4\gamma\ep$ rather than $\gamma\ep$; see \citet{bg70}.}
\begin{equation}
E_{\rm KN} \cong ({m_e c^2\over 4\times 2.7 \Theta}) \times { \Dop \over 
\Gamma(1-\beta\bar\mu_*)}\times { 1\over(1-\cos\bar\psi^\prime)}
  \cong {20 {\rm GeV} \over (1-\cos\bar\psi^\prime)}\;.
\label{EgbreakKN}
\end{equation}
The three terms making up the middle expression represent, from left
to right, the transition energy for a monochromatic radiation field
with mean photon energy $\approx 2.70 \Theta m_ec^2 $, a factor for
the dependence on jet speed, stellar location and observer direction,
and an angular factor accounting for the stellar azimuth.  The
$\phi = 0$ and $\phi = \pi $ cases in Fig.\ 2a have, roughly,
$\cos\psi^\prime \sim -0.9$ and $\cos\psi^\prime \sim +0.8$, which
accounts for the large ranges in photon energies corresponding to the
transition to the KN behavior.  KN effects are quite substantial,
however, even at lower energies than given by eq.\ (\ref{EgbreakKN}),
because of the gradual change in the Compton cross section over the
range where recoil effects start to become important.

Because the transition to the KN regime occurs at lower values of
scattered photon energy at $\phi = 0$ where the flux is higher than
for the $\phi = \pi$ case where the flux is lower, the CSSR process
will produce a softer spectra with increasing flux at multi-GeV -- TeV
energies. This is the same behavior as inferred from the effects of
$\gamma\gamma$ absorption on the emitted radiation spectrum
\citep{bd05}, and so would enhance this behavior.

\begin{figure}[t]
\vskip-0.5in
\epsscale{1.1}
\plottwo{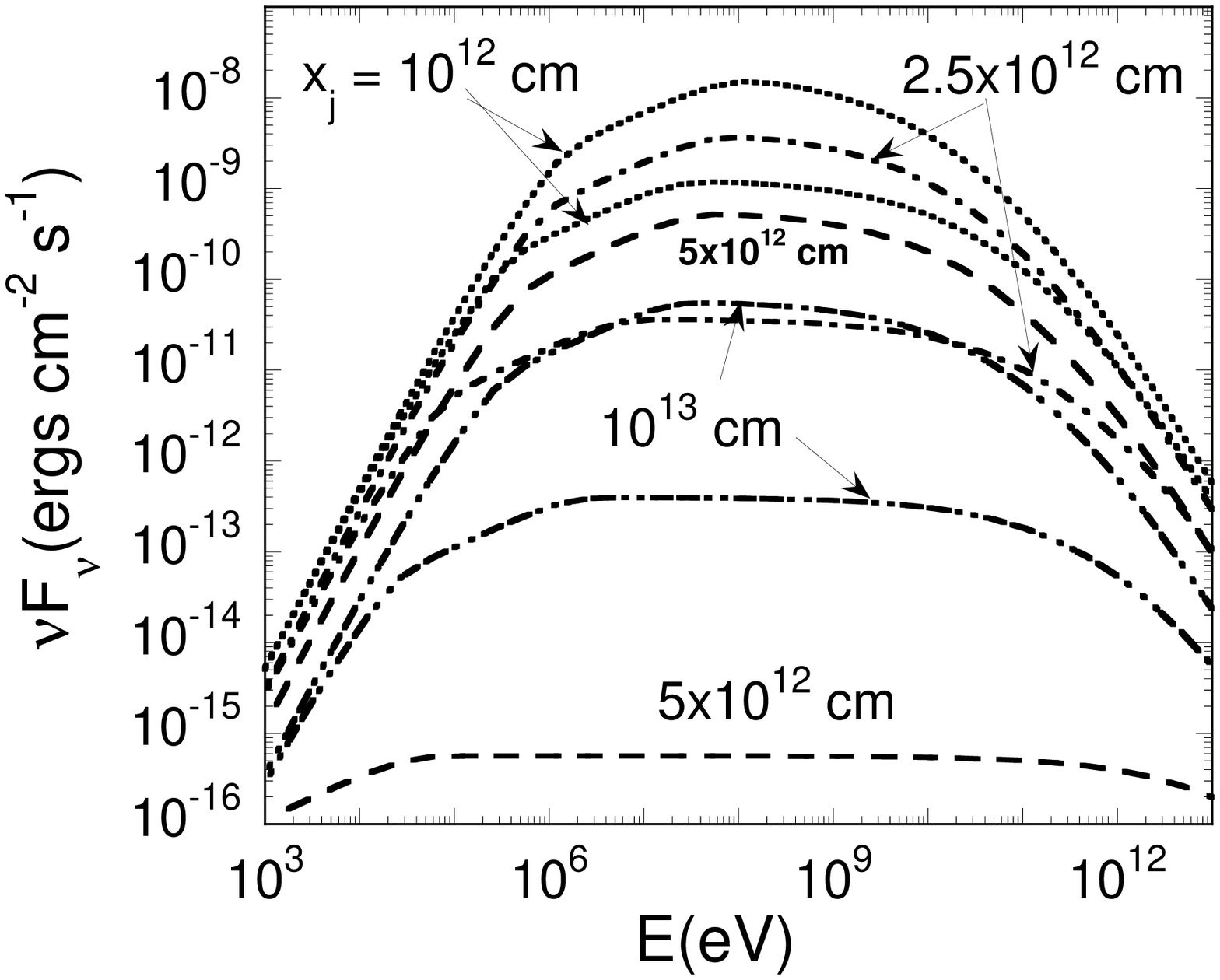}{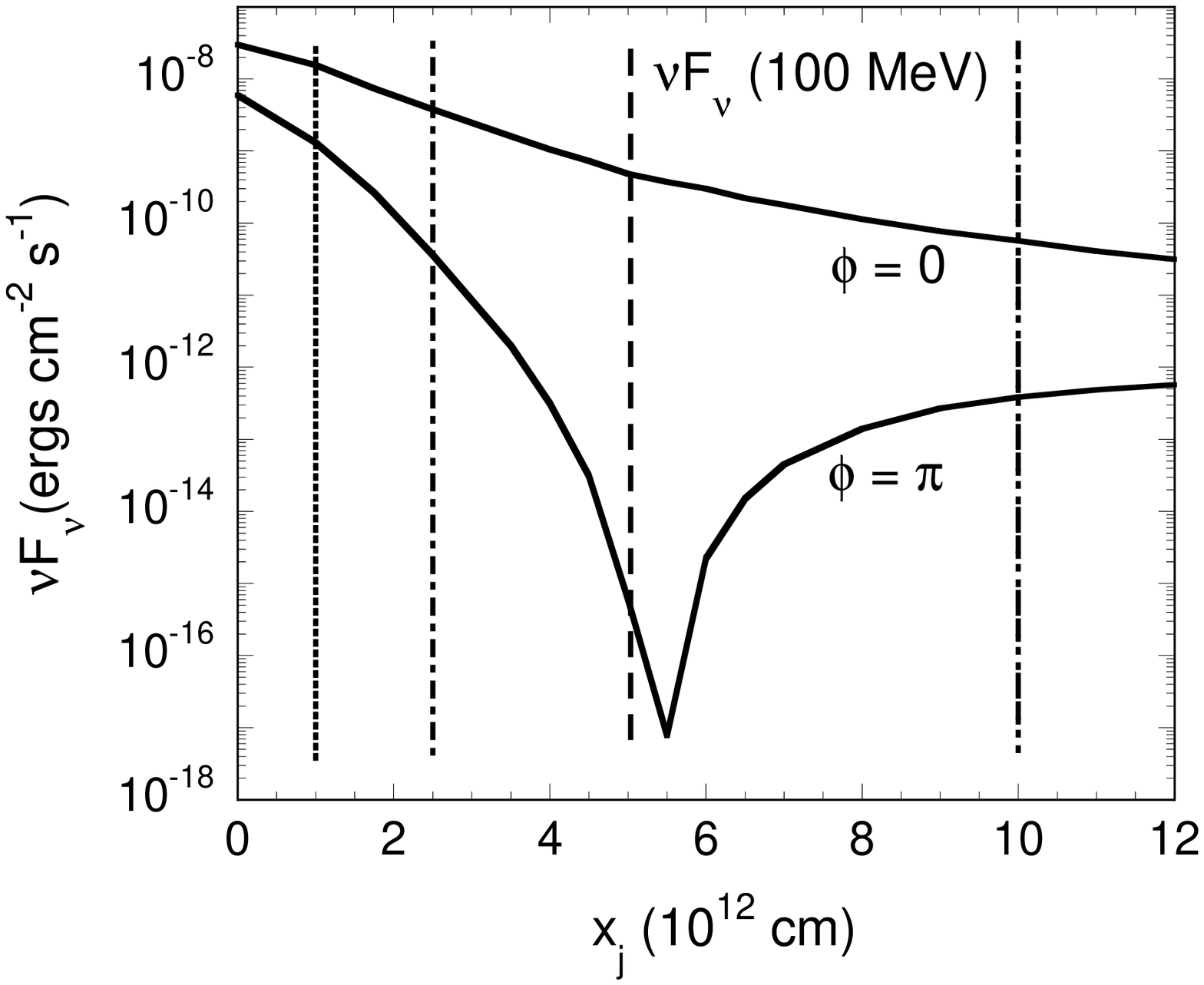}
\caption{The $\nu F_\nu$ CSSR SED
as a function of jet height $x$ at phases $\phi = 0$ and $\ \pi$, for
standard parameters given in Table 1.  (a) Dependence of the SED on
$x$ at phases $\phi = 0$ and $\phi = \pi$ shown in the upper and lower
curves, respectively, of each pair. (b) The value of $\nu F_\nu$ at 100
MeV at $\phi = 0$ and $\phi = \pi$.  Effects of $\gamma\gamma$
attenuation on the SEDs are not included here, but is unimportant
below $\approx 50$ GeV.}
\end{figure}

Besides the rapid transition to the steep KN behavior at energies
$\lesssim$ 100 GeV, Fig.\ 2a shows that the CSSR spectra are strongly
modulated in flux as a function of phase of the binary system.  Also
shown in Fig.\ 2a by the solid curve is the phase-averaged CSSR SED,
which is approximately equal to the $\phi =\pi/2$ curve (actually
closer to $\phi \cong 7/16$).  Fig.\ 2b shows calculations of CSSR at
$\phi = \pi/2$ for different values of $p$ and $x$. Due to the
diminution of the target photon density, the absolute value of the
flux decreases rapidly with jet height $x$, as expected.  The harder
electron spectrum causes the peak of the CSSR SED to shift to higher
energies, but the spectrum still falls rapidly when KN effects become
important.

Fig.\ 3 shows in more detail how the CSSR SED depends on jet height
$x$ at phases $\phi = 0$ and $\pi$.  For these parameters, the $\nu
F_\nu$ spectrum peaks between 100 MeV and 1 GeV, as shown in Fig.\ 3a.
If $\gamma$-ray emission from microquasars is due to this process,
then strong phase-dependent modulation is expected if the jet
electrons radiate on a distance scale comparable to the orbital
radius.  As seen in Fig.\ 3b, the CSSR flux nulls at $\phi = \pi$ when
the inclination angle $\theta \cong \arctan (d/x)$, which takes place at $x
\cong 5.4\times 10^{12}$ cm for our parameters. This happens because
the target photons are directed almost exactly ``tail-on" past those electrons
which would scatter photons into our observing direction. This effect
will be somewhat ameliorated in calculations where the star has a finite
extent.

Unlike the $\gamma\gamma$ attenuation process, which ceases to be
effective at producing even 10\% orbital modulation of the VHE
emission when $x \approx 10^{13}$ cm \citep{bd05}, the CSSR process
will produce significant phase-dependent modulation of the $\sim 100$
MeV -- GeV emission to much larger distances. The 100 MeV fluxes at
$\phi = 0$ and $\phi = \pi$ differ by a factor of 2 at $x \approx
5\times 10^{13}$ cm, and by a factor of 10\% at $x \approx 5\times
10^{14}$ cm. Therefore even if the $\gamma$-ray emission is not
produced in the inner jet, GLAST could still detect significant
orbital modulation of the $\gamma$ rays. The large range of distances
that produce significant orbital modulation also makes the claim
\citep{mas04} of periodic variability of the EGRET emission from 3EG
J0241+6103 associated with LSI+61$^\circ$303 more believable, and
suggests that the EGRET data for 3EG J1824-1514 should be reanalyzed
to establish strong limits to modulation at the orbital period of LS
5039.

\begin{figure}[t]
\vskip-1.5in
\epsscale{1.0}
\plotone{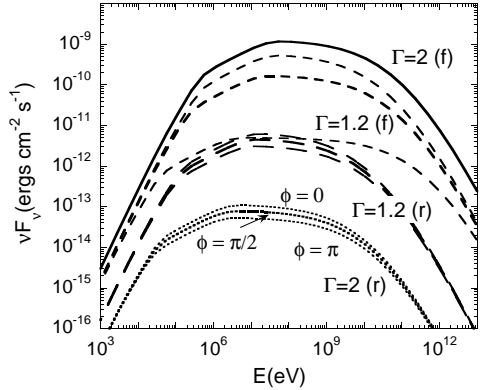}
\caption{Dependence of the CSSR SED of a 
two-sided jet on the bulk Lorentz factor $\Gamma$ of the plasma. The
SEDs for the forward (f) and reverse (r) jets are shown for $\Gamma =
1.2$ and $\Gamma = 2$ at phases $\phi = 0, \pi/2$ , and $\pi$ (see
Table 1 for parameters). Only the $\phi = \pi/2$ curve is shown for
the $\Gamma = 2$ forward jet; the other phase-dependent spectra for
this case are shown in Fig.\ 2a. }
\end{figure}

Fig.\ 4 illustrates the dependence of the fluxes of a two-sided jet
for model jets with $\Gamma = 1.2$ and 2. The forward (f) jets are
inclined at $\theta = 25^\circ$ to the observer, and the reverse (r)
jets have $\theta = 155^\circ$. Even at mildly
relativistic Lorentz factors, the Doppler boosting can lead to differences
of several orders of magnitude between the fluxes from the forward and
reverse jets. As shown in Appendix B, the beaming factors for
Thomson-scattered radiation for external photons that enter the jet
from behind and in front of the jet are $\propto
\Dop^{3+p}(1-\mu)^{(p+1)/2}$ and $\propto
\Dop^{3+p}(1+\mu)^{(p+1)/2}$, respectively. For the standard
parameters, $\mu = 0.9063$, $p = 3$, and $\beta = 0.866$ for $\Gamma =
2$. Thus the ratio of scattered fluxes of the forward and reverse jets
range from $\approx 800$ for target photons entering from behind, to
$\approx 10^8$ for target photons entering from in front of the
jet. Intermediate values are found for stellar photons that enter at
shallow angles, as calculated in Fig.\ 4.  This shows that CSSR from
the reverse jet can generally be neglected for even mildly
relativistic jets.

\section{Compton-Scattering Leptonic Jet Model for LS 5039}

In this section, we apply the preceeding results to LS 5039 data, showing that
\begin{itemize}
\item It is very difficult to fit the HESS with a CSSR model because of the strong curvature
produced by KN effects;
\item An improbable leptonic model with electrons accelerated
with maximum efficiency with a distribution displaying no spectral 
breaks is required to fit both EGRET and HESS data, assuming that these data 
are representative of the simultaneously measured spectrum of LS 5039;
\item The EGRET and HESS data can be separately fit with emission that is primarily 
from the CSSR and SSC processes, respectively, which implies significant
aperiodic variability of the $\gamma$ rays.
\end{itemize}
We furthermore consider other possible origins for the broadband emission from 
LS 5039, including hadronic and combined hadronic and leptonic models, and the 
possibility that the EGRET emission associated with LS 5039 originates
from another source in the LS 5039 field, or is contaminated by, e.g., diffuse 
radiation.

Fig.\ 5 shows non-simultaneous data \citep{aha05} taken at radio,
optical, X-ray, and $\gamma$-ray wavelengths, with EGRET data points
taken from \citet{bp04a}. The HESS data are described by a power-law
spectrum between 250 GeV and 4 TeV with photon index $=2.1\pm 0.15$
\citep{aha05}. We find that fitting the HESS spectrum with a pure CSSR
model faces severe difficulties. This is due to the strong KN decline
and marked convexity of the CSSR radiation spectrum at TeV energies.
Extremely hard $(p \approx 2)$ electron spectra extending to very high
energies are required for such a model to work, but in all cases the
CSSR SEDs display strong curvature that should be measurable with more
sensitive HESS data.

\begin{figure}
\vskip-1.0in
\epsscale{1.0}
\plotone{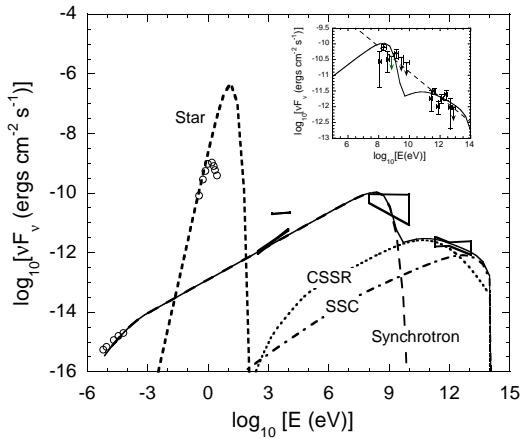}
\caption{Leptonic jet model fit for the LS 5039
data, using Model A parameters given in Table 1. 
The low and high X-ray spectra are from XMM and 
RXTE observations, respectively. The optical data have not been
dereddened. This figure illustrates the difficulty that leptonic jet
models have to fit the joint nonsimultaneous EGRET and HESS data if
these data are assumed to represent the broadband LS 5039 $\gamma$-ray
spectrum measured at a single epoch.  Inset shows the fit to the EGRET
and HESS data in more detail, with the dashed line indicating the
small range of power-law spectra that connect the EGRET and HESS
data. Total spectrum is shown by the solid curve, with components as
labeled.}
\end{figure}

Assume now that the EGRET and HESS data represent the $\gamma$-ray
spectrum of LS 5039 that would be measured at a single epoch.  In
order to fit both the EGRET and HESS data simultaneously, we find that
a CSSR fit to the EGRET data will produce too soft of a spectrum in
the HESS range. This is likewise the case for an SSC model fit to the
EGRET data. A leptonic SSC model can fit the HESS data alone, as we
show below. But an SSC model fit to the HESS data means that the EGRET
data require a separate explanation. One possibility considered in
Fig.\ 5 is that the high-energy extension of the synchrotron spectrum
could produce emission in the EGRET band. For the hard synchrotron
spectrum required in this case, it is then necessary to fine-tune the
joint CSSR and SSC spectra to obtain an acceptable fit to the joint
EGRET and HESS data (see inset of Fig.\ 5).  Model A parameters given
in Table 1 are used in this fit, noting that the mean magnetic field
$B\lesssim $ 10 G in order to have sufficiently energetic electrons to
scatter target photons to multi-TeV energies (eq.\
[\ref{gammamaxsyn}]).

It is interesting to note that the radio and low-state X-ray emission
from LS 5039 can be connected by a single power-law. This suggests
that the radio through X/$\gamma$-ray emission could originate from
the same (nonthermal jet synchrotron) process, as has been argued to
be the case for models of the LMXBs XTE J1118+480 \citep{mff01} 
and GX 339-4 \citep{mar03} on the basis of radio, IR, and X-ray spectral
correlations. In those models, a broken power-law jet synchrotron
radiation spectrum resulting from adiabatic and radiative cooling is
used to fit the data. A similar type of model, as shown in Fig.\ 5,
could in principle fit the radio/low-state X-ray data from LS 5039 if
the synchrotron cooling frequency lies either below radio frequencies
or above $\sim 100$ MeV $\gamma$-ray energies. 

Nevertheless, we do not consider this a viable explanation, as it
requires acceleration of electrons with maximal efficiency (eq.\
[\ref{gammamaxsyn}]) with no strong evidence for a synchrotron cooling
break from the radio through $\sim 100 $ MeV $\gamma$-ray regime. It
also requires fine-tuning of the CSSR and SSC spectra to fit the HESS
data. Moreover, in this jet model, as well as in the models by Markoff
et al., the radio emission begins to be self-absorbed below $\sim 10$
GHz (see Fig.\ 5) for a comoving jet radius $r_b = 2\times 10^{11}$ cm
(Table 1) needed to give the requisite SSC flux. The radio data of LS
5039, by contrast, shows no evidence for self-absorption.

Thus if the EGRET and HESS data are to be jointly fit, it seems that a
leptonic jet model does not work. Yet a simple hadronic model, where
the emission is due to a power-law spectrum of cosmic rays accelerated
by the outer parts of the jet, which then interact with the ambient
medium or the dense matter field of the stellar wind \citep{romero03}
to produce secondary nuclear pion-decay $\gamma$ rays, seems no more
likely. The dot-dashed line connecting the EGRET and HESS data in the
inset to Fig.\ 5 suggests that a single power law with number index
$\simeq 2.4$ nearly connects the two data sets.  But a power-law
cosmic-ray spectrum produces a convex $\gamma$-ray spectrum at a few
hundred MeV due to the pion production and decay processes
\citep{rco05} that would miss the EGRET data points which are most
significant.

\begin{figure}[t]
\vskip-1.0in
\epsscale{1.0}
\plotone{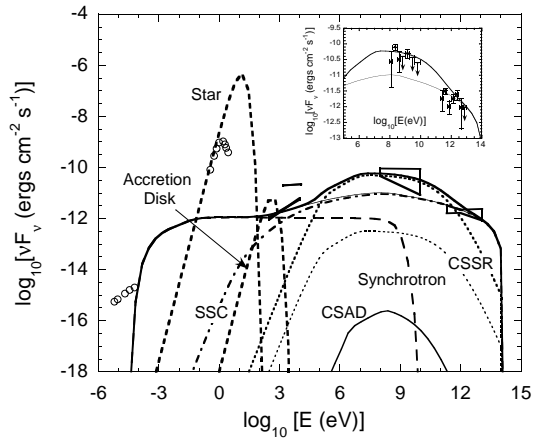}
\caption{Leptonic jet model fit for the LS 5039
data, using Model B parameters given in Table 1. 
The heavy and light solid curves in the main panel and inset gives
the total spectra for Model B parameters with $x = 10^{13}$ cm and 
$x=10^{14}$ cm, respectively. }
\end{figure}

Any leptonic jet model would predict stochastic variability, so
clearly the assumption that needs to be relaxed is that the EGRET and
HESS data, taken years apart, are typical of a contemporaneously
measured data set. Leptonic jet models must be highly variable
in view of the short radiative cooling times of nonthermal jet electrons,
which require active mechanisms for accelerating the electrons, 
such as internal shocks from variable relativistic plasma outflows 
that form the microquasar jets.

Fig.\ 6 shows an example of a fit where the EGRET
data are due primarily to the CSSR process, and the HESS data are due
primarily to the SSC process. Here we take $r_b = 10^{11}$ cm. 
The spectral components, and the heavy
solid and dotted curves for the CSSR process and the total spectrum,
respectively, refer to the case when $x = 10^{13}$ cm. The light solid
curve for the total spectrum and the light dotted curve for the CSSR
process refers to the case when $x = 10^{14}$ cm.  All other
parameters remain the same (we let $\phi = \pi/2$ in our fits).  The
difference between the two spectral fits are then due to the effects
on the CSSR component arising from the different distances of the jet
from the star. We do not fit the radio emission, which is 
produced at scales as large as $\sim 1000$ AU \citep{par00,par02}.

Also shown for completeness is a model accretion-disk spectrum with
emission that is comparable to the low-state X-ray flux, along with
the associated CSAD radiation. When $x \gg 10^{12}$ cm, the CSAD
process makes no significant contribution to the $\gamma$-ray
emission. It is surprising, however, how weak and soft (here we use an
effective temperature of 0.3 keV) the accretion-disk radiation field
has to be in comparison with the flux level of the jet radiation, and even
here the model overproduces the low-state X-ray observations. 
The use of more realistic disk models with hard X-ray emissions could
better fit the low-state X-ray data, as observed with XMM-Newton, as
well as X-ray states measured with RXTE \citep{bos05}, though these
authors favor a jet origin for the X-rays in view of the spectral
index/flux correlation and the smooth transition that is indicated
between the X-ray and $\gamma$-ray components
\citep[the X-ray high state shown
in Figs.\ 5 and 6 represent the historical 1998 high fluxes measured
with RXTE, but may be contaminated by diffuse emission; see][]{rib99,bos05}.

Fig.\ 6 shows that if the CSSR process accounts for the
origin of the $\gamma$ rays measured with EGRET, then it would exceed
the flux measured by HESS at photon energies of several hundred
GeV. This can be reversed to make a prediction that when the emission
from LS 5039 is luminous in the $\sim 100$ MeV -- GeV band, then the
HESS spectrum will be brighter and softer than reported by
\citet{aha05}.  Besides predicting stochastic variability, our model
would also predict weak or absent periodic variability of the HESS
emission on the orbital timescale associated with the SSC emission
process.  This is because SSC emission does not exhibit (for fully
isotropized electron distributions) the periodic emission signatures
of the CSSR process.  There could, however, be variability of the HESS
emission on the orbital timescale due to $\gamma\gamma$ effects if the
SSC emission originated from the inner jet, which would display the
distinctive features that we recently calculated \citep{bd05}.
Aperiodic variability of the $\gamma$ ray emission could arise from a
region as large as $\sim 10$ AU, corresponding to the size scale of
resolved, radio emission from LS 5039. The size scale of the
$\gamma$-ray emission would be difficult to constrain unless sub-hour
variability were measured with HESS or GLAST. Note that the 
$\lesssim 1^\prime$ resolution of HESS corresponds, at 2.5 kpc, to $\lesssim 100$ AU. 
EGRET localized  sources in the Milky Way's disk to $\approx 1^\circ$ at best, 
whereas GLAST will achieve a localization by a factor $\approx 10$ better than EGRET. 

The situation is, of course, more complicated, and we
can consider a number of other possibilities to explain
the broadband spectrum of LS 5039. For example,

\begin{enumerate}
\item A model
involving hadronic cascades from an inner jet source might fit the
data.  The model of \citet{aha05a} would predict strong phase-dependent
modulation of the $\gamma$-rays; if this is not detected with HESS,
this model will probably be ruled out. A more probable hadronic model, 
in accord with indications of variability, 
involves the interaction of cosmic-ray accelerated particles with the stellar
wind \citep{romero03}, though the energetics could be demanding if the cosmic rays
diffuse rapidly away from the source. 
\item A combined leptonic and hadronic model could explain
the observations. In such a model, the $\sim 100$ MeV -- GeV $\gamma$
rays are due to CSSR with a soft (cooling) electron spectrum with $p
\gtrsim 3$, and the VHE $\gamma$ rays would be formed by a hadronic
emission component due to secondary nuclear production with particles
from the surrounding medium or stellar wind, as in the model of
\citet{rco05}. In this case, we would expect variable modulated
emission at GLAST energies, and nonvarying or slowly varying radiation
at TeV energies with no periodic modulation. One would also expect
extended VHE emission around LS 5039 as the accelerated cosmic rays
diffuse away from the source. In contrast, LS 5039 emission is
reported to be consistent with a point source \citep{aha05}; even if
the extended VHE emission were too weak to detect with HESS, an
associated radio halo from pion-decay secondaries would be expected
if a dense target, such as a molecular cloud, was nearby \citep{bap05}.
\item The EGRET emission could originate from another source in the
field of LS 5039. Both PSR B1822-14 and SNR G16.8-1.1, the association
proposed by \citet{tor03}, are within the EGRET error box
\citep{aha05}. It is doubtful that a hadronic SNR component could
produce the EGRET emission without also being a significant HESS
source, so that emission from a pulsar seems more likely. Cosmic ray
irradiation of clumped cold dust not recognized in CO surveys
\citep{gct05} could also masquerade as a $\gamma$-ray source in the LS
5039 error box. Because EGRET source detection depends sensitively
on the cosmic-ray induced diffuse background, incorrect background subtraction 
for EGRET, which has a much larger spatial resolution than HESS, could
give an incorrect flux for the LS 5039 source. 
Insofar as LS 5039 is a VHE source, we would still
expect it to radiate in the $\sim 100$ MeV -- GeV band, though
possibly at a flux level lower than reported by EGRET. 
\end{enumerate} 

In spite of these possibilities, we think that emission from a
leptonic jet is the most likely explanation for the origin of the
$\gamma$-ray emission from LS 5039 and other galactic microquasars.
We agree with \citet{bp04a} that the EGRET $\gamma$ rays are likely
dominated by CSSR; indeed, a model where the EGRET emission is
primarily fit with an SSC component gives a poor fit to the HESS data
\citep{pbr05}. More likely, the VHE emission observed with HESS is
mainly SSC radiation.

\section {Summary and Conclusions}

In this work, we developed a leptonic model for the VHE emission
observed from LS 5039 with HESS, where the Compton-scattered flux
depends on observer angle $\theta$, bulk Lorentz factor $\Gamma$, and
the parameters of the binary microquasar system.  If the $\gamma$-ray
flux of a galactic microquasar is due to stellar photons that are
Compton-scattered by jet electrons, then the $\g$-ray spectrum will
exhibit variability correlated with orbital phase of the companion
star. The fractional modulation depends strongly on the location of
the jet. Phase-dependent modulation of the $\gamma$ rays can result
both from the CSSR process and $\gamma\gamma$ pair-production
absorption \citep{bd05,dub05} of jet $\gamma$ rays that interact with
photons of the high-mass star.  Periodic modulation of the VHE
emission with signatures characteristic of the $\gamma\gamma$
attenuation would provide strong evidence in favor of an inner jet
model for the origin of VHE $\gamma$ radiation from LS 5039, whereas
variability of the $\sim 100$ MeV -- GeV emission on the orbital
timescale allows the $\gamma$ rays to originate from either the inner
or extended ($d \lesssim x \lesssim 100 d$) jet.

Our study of the CSSR process for high-mass microquasars 
yielded a number of results:
\begin{enumerate}
\item The high flux states associated with the orbital modulation
of the binary microquasar system exhibit stronger spectral softening
than the low flux states as a result stronger KN effects near orbital
phase $\phi = 0$ than at phase $\phi = \pi$;
\item An expression was derived, eq.\ (\ref{EgbreakT}), 
relating breaks in the CSSR spectra for scattering in the Thomson
regime to breaks in the electron distribution. These breaks can
potentially be inferred from the synchrotron radiation spectrum, and
used to test a CSSR model.
\item Another expression, eq.\ (\ref{EgbreakKN}), 
was derived that gives the photon energy for the onset of KN effects
in CSSR spectra. The KN softening can appear at energies less than
$\sim m_e c^2/(4\times 2.7\Theta) \sim 25{\rm~GeV}/T_*({\rm eV})$,
even for mildly relativistic outflows, due to the gradual onset of the
decline in the Compton cross section.
\item Beaming factors for the CSSR process were derived in Appendix B, 
and it was shown that the principal dependence of the CSSR $\nu F_\nu$
flux $f_\e \propto \Dop^{3+p}$ (eqs.\ [\ref{feKNstar}] and
[\ref{A7}]) for an electron spectrum with index $p$.
\item Ratios of CSSR fluxes from forward
and reverse jets for a microquasar system were obtained. Even for
mildly relativistic outflows $(\Gamma \sim 1.5$) of microquasar
jets, the ratios of the CSSR fluxes from the forward and reverse jets can be
many orders of magnitude.
\item Allowed ranges of parameters related to maximum 
electron energies permitted by competing Compton losses and measured
ratios of Compton-to-synchrotron fluxes were derived in Appendices C
and D, respectively, showing that a leptonic inner jet must be, at
least, mildly relativistic.
\item Significant modulation of the CSSR $\gamma$ rays with the orbital period of the star 
can take place for emission produced in the extended jet, unlike the modulation produced
by $\gamma\gamma$ attenuation, which is only important for emission produced in the inner jet.
\end{enumerate}

From these results, we tried to construct a purely leptonic model for
the multiwavelength SED of LS 5039. The HESS data cannot be fit with
the CSSR process alone due to the strong curvature of the scattered
spectra.  If the CSSR process is used to fit the EGRET data, then it
overproduces the emission in the HESS band. Thus, it is difficult for
the CSSR and SSC processes to fit both the HESS and EGRET data
simultaneously.  To circumvent this difficulty, we considered a model,
Fig.\ 5, where the EGRET emission is due to nonthermal synchrotron
radiation emitted by an electron distribution with a very hard
spectrum $(p \lesssim 2.5)$ that extends to the radiation-reaction
limited maximum electron energy.  Not only is this electron
distribution unrealistic by displaying no cooling break, but the model
requires fine-tuning of the CSSR and SSC processes to fit the HESS
data.

Because leptonic jet models should exhibit marked aperiodic
variability, it is doubtful that the HESS and EGRET data are
representative of the $\gamma$-ray spectrum at a fixed epoch.
Abandoning this assumption makes it feasible to fit the EGRET data
with CSSR emission (Fig.\ 6) and the HESS data with SSC emission
\citep[as in the model of][]{aa99}.  This leptonic model predicts
stochastic variability of the HESS data, and enhanced emission at
several hundred GeV energies when the $\sim 100$ MeV -- GeV emission
is in a high-flux state. The level of modulation of the AGILE/GLAST
and HESS $\gamma$-ray flux on the stellar orbital period will directly
imply the location of the $\gamma$-ray emission site; in the former
case from the inner or extended jet, and for the latter case, from
within the inner jet.  Absence of periodicity of the $\gamma$-ray
emission measured with both GLAST and HESS means that the $\gamma$
rays originate from the outer ($x \gtrsim 100 d$) jet.  Confirmation of a
leptonic jet model can be made by comparing contemporaneous X-ray and
$\gamma$-ray observations with predictions of correlated variability
\citep{gbd05}.

If the jet flow is nonrelativistic, then the jet emission must be
produced in the extended or outer jet in order that electrons can be
accelerated to sufficiently high energies to emit multi-TeV $\gamma$
rays (see Appendices B and C). Thus, if the outflows forming galactic
microquasar jets are nonrelativistic throughout the full extent of the
jet, only weak periodic modulation from Compton scattering, and
essentially no periodic modulation from $\gamma\gamma$ absorption
effects, is possible.  A mildly relativistic ($\Gamma \gtrsim 2$)
leptonic jet model in the inner regions of the jet is compatible with
the acceleration of sufficiently energetic electrons to make VHE
$\gamma$ rays throughout the HESS band. If HESS data show variability
on the orbital timescale, it means that microquasars eject
relativistic flows, which must then be decelerated to nonrelativistic
speeds by the time the jets are detected with radio telescopes. Jet
deceleration can be accomplished through radiative drag on the stellar
radiation field---which makes the $\gamma$ rays---as well as via
interactions with the surrounding medium to form shocks that could
accelerate cosmic rays. In this case, the radiative opacity should
be large and a light pair jet might be required.

The most serious limitation of our study is the use of fixed electron
energy distributions. In more realistic models, power-law
distributions of nonthermal electrons are accelerated through shock
processes and are injected and evolve through radiative, adiabatic,
and cascade processes
\citep{aa99,brp05} while the jet plasma flows away from the central star. 
To treat such a system, electron energy-loss rates associated with Compton
interactions in both the Thomson and KN regimes 
need to be derived for use in an equation for electron
energy evolution
\citep[see][for an analytic treatment in the Thomson regime]{gbd05}.
Systems where Compton losses dominate synchrotron losses can introduce
unusual hardenings in the steady-state electron distribution and
synchrotron spectrum \citep{da02}, so that it is not clear under what
conditions the assumption of a power-law electron distribution is
valid. Future work will address this question.

In summary, we have developed a leptonic jet model for galactic
microquasars to fit the EGRET and HESS $\gamma$-ray data for LS
5039, complementary to the leptonic jet models developed
by \citet{bos05a}.  This model predicts aperiodic variability, and periodic
variability correlated with the orbital motion of the star that
depends on the location of the $\gamma$-ray emission site and the
speed of the jet outflow. Joint observations with AGILE, GLAST and air
Cherenkov telescopes will quickly reveal the actual contemporaneous
$\gamma$-ray spectra of high-mass microquasars, their variability properties, 
and whether a joint CSSR/SSC model, as proposed here, is correct.

\acknowledgments
We thank V.\ Bosch-Ramon for a thorough reading of the paper and many
useful suggestions.  We also thank K.\ E.\ Mitman for assistance
during the start of this project, G.\ Romero for comments and S.\ Gupta for
corrections, and G.\ Dubus and M.\ de Naurois for helpful correspondence
concerning the HESS data.  The work of C.~D.~D.\ is supported by the
Office of Naval Research and the NASA GLAST Science Investigation
DPR-S-1563-Y. The work of M.~B.\ is supported by NASA through
XMM-Newton GO grant no. NNG04GI50G and INTEGRAL theory grant
NNG05GK59G.

\appendix

\section{Derivation of CSSR Spectrum}

Quantities in the comoving jet frame are denoted by primes\footnote{No
primes are attached to the electron Lorentz factor $\gamma$, however,
as this quantity is always referred to the comoving frame here.}. The
invariance of $u^*(\es,\Os)/\es^3$ implies that the spectral energy
density of the stellar radiation field in the comoving fluid frame is
$u^\prime(\ep,\Omega^\prime) = (\ep/\es)^3 u^*_{bb}(\e_{*},\Os)$,
where $\Os = (\arccos \mu_*,\phi_*)$, $$\e_{*} = \Gamma \ep(1 + \beta
\mup) \;, \;\mu_* = (\mup +\beta)/(1+\beta\mup)\;,\;{\rm and}\;
\phi^\prime = \phi_*-\pi\;.$$
The reverse transformations are
$$\ep = \Gamma\e_*(1-\beta\mu_*)\;,\; \mup = 
(\mu_* - \beta)/(1-\beta\mu_*) \;,\;{\rm and}\; \phi_* = \phi^\prime +\pi\;.$$ 
Eq.\ (\ref{ustarbb}) becomes
\begin{equation}
u^\prime_{bb}(\ep,\Op ) = \; 
u^0_*\;{\es^{3}\;\delta(\mup -\mup_*)\delta(\phi^\prime - \bar\phi_*)
\over \Gamma (1+\beta\mup)[\exp (\es/\Theta ) - 1] }
\;,
\label{ubbprime}
\end{equation}
where 
$$\mup_* = { \bar \mu_* - \beta 
\over 1 - \beta \bar \mu_*}\;,\; \bar \phi_* = \phi_* - \pi\;.$$
This can also be written in the form of the specific spectral photon density
\begin{equation}
n^\prime_{bb}(\ep,\Op ) = \; {u^0_*\;
\over m_ec^2} \;{\es^{2}\;\delta(\mup -\mup_*)\delta(\phi^\prime - \bar\phi_*)
\over [\exp (\es/\Theta ) - 1] }
\;.
\label{nbbprime}
\end{equation}

The nonthermal electrons Compton-scatter the target stellar photons
that intercept the jet.  The comoving-frame emissivity as a function
of scattered photon energy $\e_s^\prime$, scattered direction
$\Omega_s^\prime$, and location $x$ is given by
\begin{equation} 
j^\prime(\e_s^\prime,\Omega_s^\prime;x) = 
m_ec^3 \ep_s\int_0^\infty d\ep \oint 
d\Op \int_1^\infty d\gp \oint d\Op_e  (1-\beta_e \cos \psi^\prime)
n_{bb}^\prime(\ep,\Op;x)
n_e^\prime(\gp,\Op_e) \big( {d\sigma_{\rm C} \over d\ep_s d\Op_s}\big)\;
\label{jint}
\end{equation}
\citep{dss97}. In this expression, $n_e^\prime(\g,\Op_e)d\gp d\Op_e$ 
is the differential number of electrons per unit proper volume in the
comoving frame with Lorentz factors between $\gp$ and $\gp$ + $d\gp$
that are directed into the solid angle element $d\Op_e$ in the
direction $\Op_e$, and $\psi^\prime$ is the angle between the incident
photon and electron in the comoving frame.  From Fig.\ 1b,
$$\cos\psi^\prime = \mup\mup_e + \sqrt{1-\mu^{\prime
2}}\sqrt{1-\mu_e^{\prime 2}}\cos(\phi^\prime -\phi^\prime_e)\;,$$ and
$d\sigma_{\rm C} / d\ep_s d\Op_s $ is the differential
Compton-scattering cross section. The speed of the electron is
$\beta_e c$, and $\beta_e = \sqrt{1-\g^{-2}} \approx 1$ for the
relativistic electrons of interest in this problem.

Photons are scattered within a cone of half-opening angle $\approx
1/\gp$ by electrons with $\gp \gg 1$.  For highly relativistic
electrons, we employ the Compton-scattering cross section in the
head-on approximation, where the scattered photons travel in the same
direction as the scattering electrons. Thus 
$${d\sigma_{\rm C} \over
d\ep_s d\Op_s} = {d\sigma_{\rm C} \over d\ep_s }\;\delta (\Op_s -
\Op_e)\;,$$ and
\begin{equation}
{d\sigma_{\rm C} \over d\ep_s}\;=\;
{3\sigma_{\rm T} \over 8\gp\e_i}\;\big[
 y + y^{-1} - {2\ep_s\over \gp\e_i y} 
+ \big({\ep_s\over \gp\e_i y}\big)^2\big] \; 
H\big(\ep_s;{\e_i\over 2\gp},{2\gp\e_i \over 1+2\e_i}\big)\;
\label{dsigC}
\end{equation}
\citep{j68,bg70,ds93},
where $\sigma_{\rm T}$ is the Thomson cross section, $H(x;a,b)$ is the
Heaviside function such that $H(x;a,b) = 1$ if $a\leq x\leq b$ and
$H(x;a,b) = 0$ otherwise, and 
$$y \equiv 1 - {\ep_s\over \gp}\;\;{\rm and}\;\;
\e_i \equiv \gp\ep(1-\beta_e \cos\psi^\prime)\cong
\gp\ep(1- \cos\psi^\prime)\;,$$  
where $\ep_s$ is the scattered photon energy in the comoving
frame. The term $\e_i$ gives the photon energy in the proper frame of
the electron, and defines the regime of interaction (Thomson regime
for $\e_i \ll 1$, and KN regime for $\e_i \gg 1$).

Substituting eqs.\ (\ref{nbbprime}) and (\ref{dsigC}) into eq.\
(\ref{jint}) gives, after solving the $\delta$-functions, the result
\begin{equation} 
j^\prime(\e_s^\prime,\Omega_s^\prime;x) = 
  u_*^0  c \pi r_e^2  \ep_s\;
\int_1^\infty d\gp \;{n^\prime_e(\gp,\Op_s)
\over \gptwo}\;\int_{\ep_l}^{\ep_u} {d\ep\over \ep } 
\;{\e_*^2\over \exp(\e_*/\Theta )- 1}\;\big[ y + y^{-1} - 
{2\ep_s\over \gp\bar\e_i y} 
+ \big({\ep_s\over \gp\bar\e_i y}\big)^2\big] \; ,\label{jint1}
\end{equation}
where $\bar\e_i = \gp\ep (1-\cos\bar\psi^\prime )$, 
$$\cos\bar\psi^\prime = \mup_*\mup_s - \sqrt{1-\mu_*^{\prime 2}}
\sqrt{1-\mu_s^{\prime 2}}\cos\bar\phi_*\;.$$  
The equations relating the comoving and
observer quantities are\footnote{The following results can be applied to
sources at cosmological distances by replacing $\e$ with $(1+z)\e$ in
the right-hand-sides of the subsequent expressions.}  $$\ep_s =
{(1+z)\e\over \Dop}\;\cong {\e\over \Dop }\;,\;\mup_s = {\mu -
\beta\over 1-\beta\mu}\;,\; {\rm and}\;\phi^\prime = \phi\;.$$
The limits on the
$\ep$-integral implied by the Heaviside function are $$\ep_l =
{\ep_s\over 2\gamma(\gamma - \ep_s )(1-\cos\bar\psi^\prime)}\;\;{\rm
and}\; \ep_u = {2\ep_s\over 1-\cos\bar\psi^\prime}\;.$$

The $\nu F_\nu$ spectrum resulting from Compton-scattered stellar
radiation (CSSR) is given by
\begin{equation}
f_\e^{\rm C*} = {\Dop^4\over d_L^2}\;\ep_s J^\prime (\ep_s,\Op_s )=
 {\Dop^4\over d_L^2}\;\ep_s V_b^\prime j^\prime (\ep_s,\Op_s )\;,
\label{feC}
\end{equation}
where $\Dop $ is the Doppler factor, eq.\ (\ref{Doppler}), 
 $V_b^\prime$ is the
comoving volume of the radiating region, and $d_L$ is the luminosity
distance to the source.  

For a uniform emitting region filled with an isotropic comoving
distribution of electrons, $$V_b^\prime n^\prime_e(\gp, \Op_s) =
V_b^\prime \;{n^\prime_e(\gp)\over 4\pi} ={N^\prime_e(\gp )\over 4\pi
}\;,$$ and we obtain eq.\ (\ref{feC_1}) for the CSSR spectrum.  
In this expression, we define $$a = \Gamma
(1+\beta\mup_*)\;,\;b = \gamma (1-\cos\bar\psi^\prime)\;\;{\rm
and}\;\;g = a/\Theta
\;.$$ 
The integrals are defined as $$I_i \equiv
I_i(u_1) - I_i(u_2)\;,\;i = 1,3,$$ where
\begin{equation}
I_1(u) = \int_u^\infty dx\;{x\over \exp(x) -1}
\cong \cases{\zeta(2) - u \; ,\;u \leq 1\; , & 
\cr\cr 
(1+u)\exp(-u) 
\; ,\;u\geq 1\;,
 &  \cr}
\label{I1u}
\end{equation}
$\zeta(n)$ is the Riemann zeta function ($\zeta(2) = \pi^2/6 =
1.6449\dots$), and $$u_1 = a \ep_l/\Theta\;\;{\rm and}\; u_2 = a\ep_u/\Theta\;.$$
The function $I_2$ is analytic and is given by
\begin{equation}
I_2 = \ln \big({1-e^{-u_2}\over 1-e^{-u_1}}\big)\;.
\label{I2u}
\end{equation} 
The function 
\begin{equation}
I_3(u) = \int_u^\infty dx\;{1\over x(e^{x} -1)}
\cong {e^{-u}\over u}\;.
\label{I3u}
\end{equation}
Fig.\ A1 compares the approximations for $I_1(u)$ and $I_3(u)$ with
numerical integrations.  These approximations, which introduce at most
$\approx 10$ -- 20\% errors over a narrow range, are used in
subsequent calculations.

\begin{figure}[t]
\vskip-1.0in
\epsscale{0.5}
\plotone{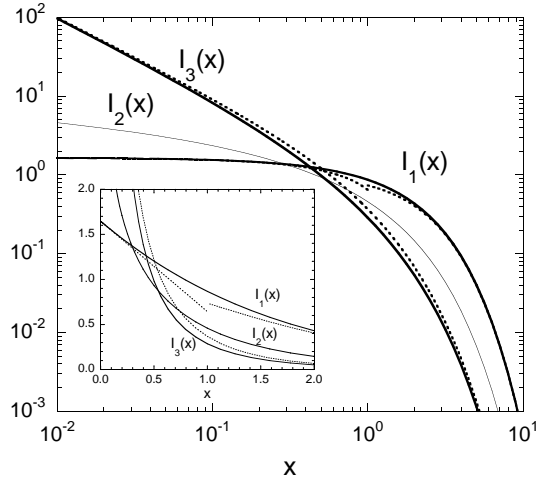}
\vskip-0.15in
\caption{Solid curves are the integrals $I_1(u)$, $I_2(u)$, and 
$I_3(u)$ given by eqs.\ (\ref{I1u}) -- (\ref{I3u}), and dotted curves
are the approximations for $I_1(u)$ and $I_3(u)$ given by
eqs. (\ref{I1u}) and (\ref{I3u}), respectively. Inset shows the
comparison on a linear scale.}
\end{figure}

\section{Comparison of Approximations for Radiation Processes}

Simpler expressions for the Compton-scattered photon spectrum from a
point source can be derived using $\delta$-function approximations for
the Thomson and KN regimes in the head-on approximation
\citep{ds93}. The accuracy of these expressions are indicated by
comparing with accurate results obtained through numerical
integrations of eqs.\ (\ref{feC_1}) and (\ref{fept}).

In the Thomson regime, we approximate the differential cross section
by the expression
\begin{equation}
{d\sigma_{\rm T}\over d\ep_s} = \sigma_{\rm T} \delta 
(\ep_s -\gamma \e_i)H(1-\e_i)\;,
\label{dsigmaTdeps}
\end{equation}
where the Heaviside function with a single argument is defined by
$H(x) = 1$ for $x \geq 0$, and $H(x) = 0$ otherwise. The Heaviside
function in eq.\ (\ref{dsigmaTdeps}) 
obviously restricts scattering to the Thomson regime.

Following the approach leading to eq.\ (\ref{feC_1}), but now using
cross section (\ref{dsigmaTdeps}) and approximating the star by a
monochromatic point source with differential energy density $$
u_*(\e_*,\Omega_*;r) = \hat u_* \delta(\e_* -2.70\Theta) \delta(\mu_*
- \bar\mu_*) \delta(\phi_* -\bar\phi_*)\;, \; \hat u_* = {L_*\over
4\pi r^2 c}\;, $$ we obtain the $\delta$-function approximation to the
CSSR $\nu F_\nu$ spectrum in the Thomson regime, given by
\begin{equation}
f_\e^{\rm T*} = \Dop^4 \;{c\sigma_{\rm T}\hat u_* (1-\beta\bar\mu_*
)^2 (1-\cos\bar\psi^\prime)^2\Gamma^2 \over 8\pi d_L^2}\;\hat\g^3
N^\prime_e (\hat \g)\; H\big[ {\Dop \over 2.70\Theta \Gamma (1-\beta
\bar \mu_* ) (1-\cos\bar\psi^\prime)} -\e \;\big]\;.
\label{feTstar}
\end{equation}
Here $$\hat \gamma = \sqrt{ {\e \over 2.70\Theta \Gamma \Dop
(1-\beta \bar \mu_* ) (1-\cos\bar\psi^\prime)}}\;,$$
$$\cos\bar\psi^\prime = \mu_*^\prime \mup_s - \sqrt{1-
\mu_*^{\prime 2} } \sqrt{1-\mu_s^{\prime 2}}
\cos (\phi^\prime - \bar\phi_* )\;,$$
and, as before, $\mu_*^\prime = (\bar\mu_* -
\beta)/(1-\beta\bar\mu_*)$ and $\mup_s = (\mu -
 \beta)/(1-\beta\mu )$. The factor
$2.70$ arises because the mean energy of a photon in a blackbody
radiation field is $[3\zeta(4)/\zeta(3)] k_{\rm B}T \cong 2.70
k_{\rm B}T$.

For a point source radiation field located behind the jet, $\bar\mu_*
\rightarrow 1$ and $\cos\bar\psi^\prime \rightarrow \mup_s$.
Replacing $2.70\Theta$ with the dimensionless soft photon energy
$\e_0$, we obtain
\begin{equation}
f_\e^{\rm T,pt} = \Dop^6 \;{\sigma_{\rm T} (1-\mu )^2 \over 32 \pi^2
x^2 d_L^2}\;
\int_0^{1/\e(1-\mu)}d\e_o\; L(\e_0) \g_{\rm T}^3 N^\prime_e(\g_{\rm T})\;
\label{feT}
\end{equation}
\citep{dsm92}, where 
$$\g_{\rm T} = {1\over \Dop}\sqrt{\e\over \e_0(1-\mu)}\;.$$ This
result is written in a form suitable for integration over the spectrum
of an isotropic point source with spectral luminosity $L(\e_0)$.  

The beaming properties of the scattered spectrum are easily derived from
eq.\ (\ref{feTstar}). If $N_e^\prime(\g ) \propto \g^{-p}$, then
\begin{equation}
f_\e^{\rm T*} \propto
\Dop^{(5+p)/2}[\Gamma(1-\beta\bar\mu_*)
(1-\cos\bar \psi^\prime )]^{(p+1)/2}(\e/\e_0)^{(3-p)/2}
\;.
\label{beaming1}
\end{equation}
For a point source behind the jet, $\bar\mu_* 
\rightarrow 1$, and $[\dots ] \rightarrow \Dop(1-\mu)$
in eq.\ (\ref{beaming1}), so that
$$f_\e^{\rm T,pt} \propto
\Dop^{3+p}(1-\mu)^{(p+1)/2}(\e/\e_0)^{(3-p)/2}
\;.$$
For a point source in front of the jet, $\bar\mu_* 
\rightarrow -1$, and $[\dots ] \rightarrow \Dop(1+\mu)$
in eq.\ (\ref{beaming1}), so that
$$f_\e^{\rm T,ptf} \propto
\Dop^{3+p}(1+\mu)^{(p+1)/2}(\e/\e_0)^{(3-p)/2}
\;,$$
which has the same beaming dependence as an external isotropic photon
field \citep{der95}.  Because these results span the possible
locations of the stellar point source radiation field, we see that the
principal dependence of the beaming factor in the Thomson regime is
$f_\e^{\rm T*} \propto
\Dop^{3+p}(\e/\e_0)^{(3-p)/2}$ multiplied by an angle-dependent factor; 
the exact beaming factor 
for the CSSR process in the Thomson regime is given by eq.\
(\ref{beaming1}).
 
In the KN regime, we use the $\delta$-function approximation
\begin{equation}
{d\sigma_{\rm KN}\over d\ep_s} = {3\sigma_{\rm T}\over
8\e_i}\ln(2e^{1/2}\e_i) \delta (\ep_s -\gamma)H(\e_i-1 )\;
\label{dsigmaKNdeps}
\end{equation}
\citep{ds93,lw04} for the cross section. 
The CSSR $\nu F_\nu$ spectrum of point-source emission
scattered by a jet in the KN regime is given for approximation
(\ref{dsigmaKNdeps}) by the expression $$f_\e^{KN*} = \Dop^6\;{3 c
\sigma_{\rm T} \hat u_* \g_{\rm KN}^3 N_e^\prime (\g_{\rm KN})
\over 32 \pi d_L^2 (2.70\Theta )^2 \e^2}\;\ln[2e^{1/2} 
\g_{\rm KN} \cdot 2.70\Theta \Gamma (1-\beta \bar\mu_*)(1-
\cos \bar\psi^\prime ) ]\times$$
\begin{equation}
\;H\big[ \e - {\Dop \over 2.70\Theta \Gamma (1-\beta \bar \mu_* ) 
(1-\cos\bar\psi^\prime)} \;\big]\;,
\label{feKNstar}
\end{equation}
where
\begin{equation}
\gamma_{KN} = {\e \over \Dop}\;.
\label{gammaKN}
\end{equation}

The KN regime $\nu F_\nu$ spectrum of an isotropically emitting
point-source of radiation located behind the jet is given by
the expression
\begin{equation}
f_\e^{\rm KN} = \Dop^6 \;{ 3 \sigma_{\rm T} \g_{\rm KN}^3 
N^\prime_e(\g_{\rm KN})\over 128 \pi^2 x^2 d_L^2 \e^2}\;
\int_{1/\e(1-\mu)}^\infty d\e_o\; {L(\e_0)\over \e_0^2} 
\ln[2e^{1/2}\e\e_0(1-\mu)]\;.
\label{feKN}
\end{equation}
The principal dependence of the beaming factor in the KN regime for a
point source behind $(-)$ and in front of $(+)$ of the 
jet, from eqs.\ (\ref{feKNstar}) and (\ref{gammaKN}) and the 
results following eq.\ (\ref{beaming1}), goes as 
\begin{equation}
f_\e^{\rm KN} \propto\Dop^{3+p}\ln[\e\e_0(1\pm\mu)]\e^{1-p}\;.
\label{A7}
\end{equation}
Because of the slowly varying logarithmic factor, the Doppler 
dependence of the beaming
factor in the KN regime goes as $f_\e^{\rm KN} \propto\Dop^{3+p}$, 
as in the case for a surrounding isotropic external radiation field
\citep{gkm01}.

\begin{figure}[t]
\vskip-1.5in
\epsscale{0.6}
\plotone{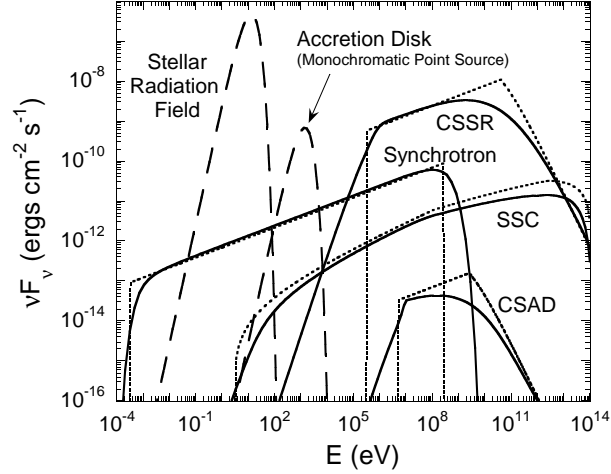}
\vskip-0.25in
\caption{Spectra of stellar and accretion-disk radiation 
fields (long-dashed curves), and accurate and approximate calculations
of the spectral components emitted by a one-sided microquasar jet. The
jet spectral components include the synchrotron spectrum, the
synchrotron self-Compton (SSC) spectrum, the Compton-scattered stellar
radiation (CSSR) spectrum, and the Compton-scattered accretion disk
(CSAD) spectrum. Accurate calculations for the synchrotron spectrum,
the SSC spectrum, and the CSSR and CSAD spectra in the head-on
approximation are shown by the solid curves, whereas $\delta$-function
approximations are shown by the dotted lines and curves.  The
parameters of the calculation are given by the parameter study values
in Table 1, except that here the electron spectrum is a single power
law with spectral index $p = 2.5$ for $10^2 \leq \gamma \leq 10^8$,
and $\phi = 0$.}
\end{figure}

Fig.\ B1 gives a comparison of the $\delta$-function (dotted lines)
approximations and the accurate (solid curves) calculations of the
CSSR and CSAD processes, respectively. The lower and upper branches of
the approximations for the CSSR and CSAD processes represent the
Thomson and KN $\delta$-function approximations, respectively. Here we
treat a monochromatic point source of radiation that radiates
isotropically from behind a one-sided jet with Lorentz factor $\Gamma
= 2$. The parameters of the stellar radiation field and accretion disk
point source are given by Table 1 parameter study values, though here
we use only a single power-law electron distribution between electron
Lorentz factors $\gamma = 10^2$ and $10^8$ with $p = 2.5$.  The
accretion disk luminosity is $10^{36}$ ergs s$^{-1}$, and the
accretion disk is assumed to emit monochromatically at 1 keV. The
emission from this source is plotted as a thermal emitter with
effective temperature $\Theta_{pt} = 1.0/(2.70\times 511)$.  Depending
on the accuracy desired, the $\delta$-function approximations may be
adequate for calculating the spectrum of a point-source radiation
field, and is simpler when integrating over non-monochromatic
accretion-disk radiation spectra.

It is also useful to compare simple $\delta$-function approximations
for the synchrotron and SSC processes with more accurate
calculations. A $\delta$-function approximation for the synchrotron
spectrum is
\begin{equation}
f^{\rm{syn}}_\epsilon \cong {\Dop^4 \over 6\pi d_L^2} c \sigma_T u_B^\prime 
\gamma_s^3 N_e\pr(\gamma_s)\;,\; \g_s = \sqrt{\ep\over \Dop\e_B}\;,
\label{f_sync}
\end{equation}
where $\e_B = B{\rm(G)}/4.414\times 10^{13}{\rm~G}$, and $B$(G) is the
magnetic field in the comoving frame. A $\delta$-function
approximation for the SSC spectrum is
\begin{equation}
f^{\rm{SSC}}_\epsilon \cong \Dop^4 \;{c \sigma_T^2 r_b u_B^\prime\over
12\pi d_L^2 V^\prime_b}\;\big({\ep\over \e_B}\big)^{3/2}
\int_0^{\rm{min}(\ep,1/\ep)} d\ep_i \;\e_i^{\prime -1} N^\prime_e(
\sqrt{\ep \over \ep_i}) N^\prime_e( \sqrt{\ep_i \over \epsilon_B})
\label{f_SSC}
\end{equation}
\citep{ds02}, where $u_B^\prime = B^2/8\pi $ and $V_b^\prime = 4\pi r_b^3/3$. 
The only provision made in this approximation to account for scattering in the KN
regime is to terminate the scattering when $\gamma\ep_i \geq 1$.

For the broken power-law electron distribution, eq.\ (\ref{elecbpl}), 
eq.\ (\ref{f_SSC}) gives
\begin{equation}
f^{\rm{SSC}}_\epsilon \cong \Dop^4 \;{c \sigma_T^2 u_B^\prime\over 16\pi^2
d_L^2 r_b^2}
\;K^2\big({\ep\over \e_B}\big)^{3/2} 
\big[ \g_1^{2(q-p)} {\cal I}_1 + \g_1^{q-p}({\cal I}_2 + 
{\cal I}_3 ) + {\cal I}_4 \big]\;,
\label{f_SSC_pl}
\end{equation}
where $$ {\cal I}_1 = \big({\ep\over \e_B}\big)^{-q/2}
\ln \big[ {\min(\e^{\prime -1},\ep/\g_0^2,\e_B\g_1^2 )\over
 \max(\ep/\g_1^2, \e_B\g_0^2 )} \big]\;,$$
$$ {\cal I}_2 = {2\e_B^{q/2} \e^{\prime -p/2}\over p-q}\;
\{ \big[ {\min(\e^{\prime -1},\ep/\g_1^2,\e_B\g_1^2 \big]^{(p-q)/2}
-  \big[ \max(\ep/\g_2^2, \e_B\g_0^2 )} \big]^{(p-q)/2}\}\;,$$
$$ {\cal I}_3 = {2\e_B^{p/2} \e^{\prime -q/2}\over q-p}\;
\{ \big[ {\min(\e^{\prime -1},\ep/\g_0^2, \e_B\g_2^2\big]^{(q-p)/2}
-  \big[ \max(\ep/\g_1^2, \e_B\g_1^2 )} \big]^{(q-p)/2}\}\;,\; {\rm and}$$
$$ {\cal I}_4 = \big({\ep\over \e_B}\big)^{-p/2}  
\ln \big[ {\min(\e^{\prime -1},\ep/\g_1^2,\e_B\g_2^2 )\over 
\max(\ep/\g_2^2, \e_B\g_1^2 )} \big]\;$$
\citep[see also][]{tave98}. In deriving this expression, we 
assume that the mean escape time of a synchrotron photon from 
the uniform spherical emitting region is $3r_b/4c$. 

The quality of the $\delta$-function approximations for the synchrotron and
SSC processes is shown in Fig.\ B1. Note that synchrotron self-absorption plays
a role at the lowest synchrotron photon energies, and is not considered in the 
approximate curve.

\section{Allowed Parameters for Leptonic Jet Model from Radiation Reaction Limit}

We first generalize the well-known expression for the maximum electron
Lorentz factor obtained by comparing the optimal Fermi acceleration
rate with the synchrotron energy-loss rate \citep{gfr83,rm98} to
include Compton losses on the CSSR field.  Besides treating KN effects
on the energy-loss rates \citep{aha05a}, our results also correct for
relativistic jet motions on the Doppler-boosted stellar radiation
field.

The electron gyration frequency $\omega_g = eB/m_e c \gamma $ in a
magnetic field with mean intensity $B$. Note a number of underlying
assumptions in this treatment: that the mean magnetic field is uniform
throughout the emitting volume, and that it is randomly oriented. The
acceleration rate $\omega_{acc}$ in Fermi processes cannot exceed the
gyration frequency because an electron gains at most a fraction of its
energy when executing a single gyration in first-order shock or
second-order stochastic Fermi processes. Thus $\omega_{acc}= \eta
\omega_g$, with $\eta \lesssim 1$. The fractional synchrotron
energy-loss rate $\omega_{syn} = |-\dot\gamma_{syn}/\gamma | =
\sigma_{\rm T} B^2\gamma/6\pi m_e c$ which, when equated with the
acceleration rate, gives
\begin{equation}
\gamma_{max} = \eta^{1/2} \sqrt{6\pi e\over \sigma_{\rm T} B} 
\cong {1.2\times 10^8 \eta^{1/2}\over
\sqrt{B({\rm G})} } \;,
\label{gammamaxsyn}
\end{equation}
and a maximum synchrotron frequency of $h\nu_{max} \cong 
\Dop(\hbar eB/m_e c)\gamma^2/(1+z) \cong 160 \eta \Dop/(1+z)
{\rm ~MeV}$.  

Scattering in the Klein-Nishina regime is important when $4\gamma\ep =
4\gamma[2.70\Theta\Gamma(1-\beta\bar\mu_*)] \gtrsim 1$ (compare eq.\
[\ref{EgbreakKN}]), that is, when $\gamma \gtrsim 5\times 10^4/[k_{\rm
B}T_*({\rm eV})\Gamma(1-\beta\bar\mu_*)]$.  The fractional energy-loss
rate in the extreme KN regime for a blackbody radiation field with
dimensionless temperature $\Theta$ is
\begin{equation}
\omega_{\rm KN} = |{-\dot\gamma_{\rm KN}\over \gamma} |
 \cong {c\sigma_{\rm T}\over 16 \lC^3}
\;{\Theta^2\over \gamma}\; \ln(0.552 \gamma\Theta)\;
\label{omegaKN}
\end{equation}
\citep{bg70}, where $\lC = \lambda_{\rm C}/2\pi = 
\hbar/m_e c = 3.86\times 10^{-11}$ cm is (also called) the electron
 Compton wavelength.

The energy density $u_*^\prime$ of a stellar blackbody radiation field
in the comoving jet frame has to be corrected by a graybody factor
that accounts for (1) the dilution of the radiation field due to the
distance of the jet from the star, and (2) the reduction in energy
density due to the bulk relativistic motion of the jet. Recalling that
the energy density at the surface of a blackbody is $1/4$ times the
energy density in the interior of a blackbody cavity, the relation
between the energy densities is therefore given by
\begin{equation}
u_*^\prime = {1\over 4} \big({R_*\over r}\big)^2  
\;\Gamma^2 (1-\beta\bar\mu_* )^2 u_{bb} (\Theta )\;
\label{ustarprime}
\end{equation}
\cite[e.g.,][]{ds02}, where 
$u_{bb}(\Theta ) = {\pi^2 m_ec^2\Theta^4/ 15\lC^3}\;$
is the energy density of a blackbody stellar radiation field with
dimensionless temperature $\Theta $.

The graybody factor from eq.\ (\ref{ustarprime}) therefore gives the
maximum electron Lorentz factor $\gamma_{max}$ by solving the equation
\begin{equation}
\eta\;{eB\over m_e c }  = {\sigma_{\rm T} 
B^2\gamma^2_{max}\over 6\pi m_e c} + 
\big({R_*\over r}\big)^2  \;\Gamma^2 
(1-\beta\bar\mu_* )^2{c\sigma_{\rm T}\over 64  \lC^3}
\;\Theta^2\; \ln[0.552 \gamma_{max}\Gamma\Theta(1-\beta\bar\mu_* )]\;.
\label{gammamaxKN}
\end{equation}
which holds when $\gamma_{max}\gg [\Gamma\Theta(1-\beta\bar\mu_*
)]^{-1}$. The maximum synchrotron energy in the comoving frame is
$h\nu_{max,syn}^\prime \cong (\hbar e B/ m_e c) \gamma_{max}^2$. Eq.\
(\ref{gammamaxKN}) can be rewritten as
\begin{equation}
\gamma_{max}^2 = \eta\; {9B_{cr}\over 4\alpha_f B}\; -
 {3\pi\alpha_f\over 32}\;({B_{cr}\over B})^2
\big[ {R_* \Gamma (1-\beta\bar\mu_* )\Theta\over r}\big]^2 
\ln[0.552 \gamma_{max}\Gamma\Theta(1-\beta\bar\mu_* )]\;,
\label{gammamaxKN_1}
\end{equation}
where $\alpha_f = 1/137$ is the fine structure constant and $B_{cr} =
m_e^2c^3/e\hbar = 4.41\times 10^{13}$ G is the critical magnetic
field. Again, this expression holds as long as the argument of the
logarithm is much greater than unity.

Because of the similar dependence of the acceleration rate and the
fractional KN energy-loss rate found in eq.\ (\ref{gammamaxKN}),
Compton-scattering in the KN regime will be unimportant to limit
electron acceleration when
\begin{equation}
{r\over R_*}\; \gtrsim \alpha_f \Theta \;
\sqrt{{\pi B_{cr}\over 24\eta B} }\cong 
0.07\;{ T_*({\rm eV})\;\Gamma(1-\beta\bar\mu_* )
\over \sqrt{\eta B({\rm G})}}\;; 
\label{rKNmax}
\end{equation}
here we have taken the square root of the logarithmic term $\approx
2$.

A limit on the range of values of $B$ can be obtained by noting that
the emission of photons with energy $E_\gamma$ requires electrons with
Lorentz factors $$
\gamma  \; \gtrsim \; {E_\gamma\over \Dop m_ec^2}\;
\cong\; {2\times 10^7 E_{10}\over \Dop } \;,$$
where $E_{10}\equiv E_\gamma / 10{\rm ~TeV}$. 
From Eq.\ (\ref{gammamaxsyn}), 
\begin{equation}
B({\rm G}) \lesssim 38\eta \Dop^2/E_{10}^2\;.
\label{gammae10}
\end{equation}
With eq.\ (\ref{rKNmax}), this implies an allowed range of magnetic
fields, given by
\begin{equation}
{5\times 10^{-3}\over \eta} \; T_*^2({\rm eV})\;
\big[ {R_* \Gamma (1-\beta\bar\mu_* )\over r}\big]^2
\lesssim B({\rm G})\lesssim {38\eta \Dop^2 \over E_{10}^2}\;,
\label{Brange}
\end{equation}
for a leptonic jet model to apply to the observed VHE emission from a
microquasar.

For the parameters of LS 5039 given in Table 1,
\begin{equation}
{0.024\over \eta}{[\Gamma(1-\beta\bar\mu_*)]^2\over  r^2_{12}}
\; \lesssim \; B({\rm G})\; \lesssim\; 
{38\eta \Dop^2 \over E_{10}^2}\;,
\label{Brange2}
\end{equation}
where $r_{12} = r/10^{12}$ cm and $\bar\mu_* = \sqrt{1-d^2/r^2}$.
For a nonrelativistic outflow, as treated by \citet{aha05a}, 
$\eta \rightarrow \eta_0 \beta^2 = (0.2)^2\eta_0(\beta/0.2)^2$,
with $\eta_0 \lesssim 1$. The allowed range becomes 
\begin{equation}
{0.6\over \eta_0 r_{12}^2 (\beta/0.2)^2}
\; \lesssim \; B({\rm G})\; \lesssim\; 
{ 1.4\eta_0  (\beta/0.2)^2 \over E^2_{10}}\;\;.
\label{Brange1}
\end{equation}
If the jet is nonrelativistic, eq.\ (\ref{Brange1}) circumscribes the
mean comoving magnetic field of a leptonic jet model for LS 5039 to a
narrow range of values unless $r_{12} \gg 1$ even for the optimistic
case of $\eta_0 \cong 1$, in which case the orbital modulation of the
$\gamma$-ray emission due to $\gamma\gamma$ attenuation and, to a
lesser extent, the CSSR process will be small. Even a mildly
relativistic jet ($\Gamma \gtrsim 1.5$) opens up a much larger
allowable parameter space to permit the source of emission to
originate from the inner jet. Note that the inferred jet speed from
radio observations \citep{par02} of LS 5039 may not be representative
of the jet speed near the source due to bulk deceleration caused, for
example, by radiative drag.

\section{Allowed Parameters for Leptonic Jet Model 
from Ratio of Compton to Synchrotron Fluxes}

A further restriction on parameters for a leptonic microquasar jet
model can be obtained by noting that the ratio of the peak $\nu F_\nu$
fluxes for the Compton and synchrotron components, denoted by $\rho$,
is related to the ratio of the comoving photon $u^\prime_*$ and
magnetic field $u_B^\prime $ energy densities by the relation
\begin{equation}
\rho \equiv {f_{\e_{pk,{\rm C}}}^{\rm C}\over f_{\e_{pk,{\rm s}}}^{\rm s} } 
\approx {u^\prime_*\over u^\prime_B}\;
\label{rho}
\end{equation}
\cite[compare][for blazars]{sik97}. This expression, which neglects angular 
scattering effects, holds for scattering in the Thomson regime and so
would best be applied to emission observed in the GLAST band.  Taking
$$u^\prime_* = {L_*\over 4\pi r^2 c} \Gamma^2
(1-\beta\bar\mu_*)^2\;,$$ (compare eq.\ [\ref{ustarprime}];
$\sigma_{\rm SB} = \pi^2 m_e c^3 k_{\rm B}^4/[60\lC^3
(m_e c^2)^4 ]$ ), we find
\begin{equation}
r \; \gtrsim \; {E_{10}^2 \over 38\eta \Dop^2}\;({2L_*\over
c\rho})^{1/2} \Gamma (1-\beta\bar\mu_* )\;.
\label{secondrrange}
\end{equation}
using eq.\ (\ref{gammae10}). For the parameters of LS 5039,
\begin{equation}
r_{12} \; \gtrsim \; {0.6 E_{10}^2 \over \eta
\sqrt{\rho/100}}\;{\Gamma (1-\beta\bar\mu_* )\over \Dop^2}\;
\rightarrow\;
{15 E_{10}^2 \over \eta_0 (\beta/0.2)^2\sqrt{\rho/100}}\;,
\label{thirdrrange}
\end{equation}
where the last expression holds for a nonrelativistic
outflow. Application of this constraint implies that we properly
identify the peak nonthermal jet synchrotron flux, which is not
necessarily obvious from the broadband data (see Fig.\ 5).  Choosing a
value $\rho \approx 10^2$ suggested by Fig.\ 6 data, we conclude that
emission from the inner jet is possible for a mildly relativistic
leptonic jet model, but not for a nonrelativistic jet at the base of
the microquasar.



\begin{table}
\begin{center}
\caption{Standard values used in parameter study and model fits to LS 5039}
\vskip0.1in
\begin{tabular}{ccccc}
\hline
Quantity & Parameter & Model A &  Model B &\\ 
         & Study     &  &   &\\ 
\hline
\hline
$d_L({\rm kpc})$ & 3.0 &  &  &\\
$ T_*({\rm K})$  & 39000 &    &  \\
$ L_*({\rm ergs~s}^{-1})$  & $7\times 10^{38}$ &   &   & \\
$ R_*$(cm)  & $6.5\times 10^{11}$ &   &   & \\
$ d({\rm cm})$  & $2.5\times 10^{12}$ &   &    &\\
$\theta$ (incl.)  & 25$^\circ$ &    &    &  \\
$\Gamma$ & 2.0  &  &  &  \\
$ \Dop$  & 2.32 &   &   & \\
$B({\rm G})$ & 1.0 & 0.8 & 0.8 &  \\
$r_b ({\rm cm}) $ & $10^{11}$ &  $2\times 10^{11}$  & $ 10^{11}$  & \\ 
$W_e^\prime ({\rm ergs}) $ & $10^{38}$ &   $2\times 10^{37}$ &  $2\times 10^{38}$ &\\ 
$p$ & 3.0 &  2.25 &  3.0 &\\
$q$ & $2.0$& 1.25  &  2.0  &\\
$\g_0$  & 100 & 10 & 300\\
$\g_1$  & $10^3$ & 100 & $10^3$ \\
$ x({\rm cm})$  & $2.5\times 10^{12}$ & $2\times 10^{13}$  & $10^{13},10^{14}$  & \\
$ L_{pt}({\rm ergs~s}^{-1})$  & $ 10^{36}$ &  -- &  $ 10^{34}$  & \\
$ T_{pt}({\rm keV})$  & 1/2.70 &  -- & 0.3/2.70 & \\
\hline
\end{tabular}
\label{tab:invariant_cross_section_constants}
\end{center}
\end{table}


\begin{thebibliography}{}

\bibitem[Aharonian et al.(2005)]{aha05}Aharonian, F., et al.,
2005, Science, 309, 746

\bibitem[Aharonian et al.(2005a)]{aha05a} Aharonian, F., Anchordoqui, L.\ A., Khangulyan, D.,  
\& Montaruli, T., 2005a, astro-ph/0508658

\bibitem[Atoyan \& Aharonian(1999)]{aa99} Atoyan, A.~M., \& 
Aharonian, F.~A.\ 1999, \mnras, 302, 253 

\bibitem[Bednarek(1997)]{bed97} Bednarek, W.\ 1997, \aap,  322, 523 

\bibitem[Begelman \& Sikora(1987)]{bs87} Begelman, M.~C., \& 
Sikora, M.\ 1987, \apj, 322, 650 


\bibitem[Blumenthal \& Gould(1970)]{bg70} Blumenthal, G.~R., 
\& Gould, R.~J.\ 1970, Reviews of Modern Physics, 42, 237 


\bibitem[B\"ottcher \& Dermer(2005)]{bd05} B\"ottcher, M., \& Dermer, C.~D.\ 2005, ApJ Letters, 634, L81

\bibitem[Bosch-Ramon et al.(2005)]{bos05} Bosch-Ramon, V., 
Paredes, J.~M., Rib{\' o}, M., Miller, J.~M., Reig, P., \& Mart{\'{\i}}, 
J.\ 2005, \apj, 628, 388 

\bibitem[Bosch-Ramon et al.(2005a)]{bos05a} Bosch-Ramon, V., 
Romero, G.~E., \& Paredes, J.~M.\ 2005a, \aap, 429, 267 

\bibitem[Bosch-Ramon et al.(2005b)]{brp05} Bosch-Ramon V., Romero, 
G.\ E., \& Paredes, J.\ M.\ 2005b,  A\&A, in press (astro-ph/0509086)

\bibitem[Bosch-Ramon et al.(2005c)]{bap05} Bosch-Ramon, V., 
Aharonian, F.~A., \& Paredes, J.~M.\ 2005c, \aap, 432, 609 

\bibitem[Bosch-Ramon \& Paredes(2004a)]{bp04a}Bosch-Ramon, V., \& Paredes,
J. M., 2004a, A\&A, 417, 1075

\bibitem[Bosch-Ramon \& Paredes(2004b)]{bp04b} Bosch-Ramon, 
V., \& Paredes, J.~M.\ 2004b, \aap, 425, 1069 

\bibitem[Butt et al.(2003)]{but03} Butt, Y.~M., Maccarone, 
T.~J., \& Prantzos, N.\ 2003, \apj, 587, 748 

\bibitem[Casares et al.(2005)]{cas05}Casares, J., Rib\'o, M.,
Ribas, I., Paredes, J. M., Mart\'i, J., \& Herrero, A., 2005, MNRAS,
364, 899


\bibitem[Dermer(1995)]{der95} Dermer, C.~D.\ 1995, \apjl, 446, L63 

\bibitem[Dermer et al.(1992)]{dsm92} Dermer, C.~D., 
Schlickeiser, R., \& Mastichiadis, A.\ 1992, \aap, 256, L27 

\bibitem[Dermer \& Schlickeiser(1993)]{ds93} Dermer, 
C.~D.~\& Schlickeiser, R.\ 1993, \apj, 416, 458

\bibitem[Dermer et al.(1997)]{dss97} 
Dermer, C.~D., Sturner, S.~J., \& Schlickeiser, R.\ 1997, \apjs, 109, 103 

\bibitem[Dermer \& Schlickeiser(2002)]{ds02} Dermer, C.~D., 
\& Schlickeiser, R.\ 2002, \apj, 575, 667 

\bibitem[Dermer \& Atoyan(2002)]{da02} Dermer, C.~D., \& 
Atoyan, A.~M.\ 2002, \apjl, 568, L81 

\bibitem[Dubus(2005)] {dub05} Dubus, G., 2005, A\&A, submitted (astro-ph/0509633)

\bibitem[Fender \& Maccarone(2004)]{fm04} Fender, R., \& 
Maccarone, T.\ 2004, ASSL Vol.~304: Cosmic Gamma-Ray Sources,  
K.S. Cheng and G.E. Romero (eds.), (Kluwer: Dordrecht), p. 205 (astro-ph/0310538)

\bibitem[Georganopoulos et al.(2002)]{gak02} 
Georganopoulos, M., Aharonian, F.~A., \& Kirk, J.~G.\ 2002, \aap, 388, L25 

\bibitem[Georganopoulos et al.(2001)]{gkm01} Georganopoulos, 
M., Kirk, J.~G., \& Mastichiadis, A.\ 2001, \apj, 561, 111; 
(e) 2004, \apj, 604, 479 

\bibitem[Gregory \& Taylor(1978)]{gt78}Gregory, P. C., \& Taylor, A. R.,
1978, Nature, 272, 704

\bibitem[Grenier et al.(2005b)]{gct05} Grenier, I.~A., 
Casandjian, J.-M., \& Terrier, R.\ 2005a, Science, 307, 1292 

\bibitem[Grenier et al.(2005a)]{gre05} Grenier, I.~A., 
Bernad{\'o}, M.~M.~K., \& Romero, G.~E.\ 2005a, \apss, 297, 109 

\bibitem[Guilbert et al.(1983)]{gfr83} Guilbert, P.~W., 
Fabian, A.~C., \& Rees, M.~J.\ 1983, \mnras, 205, 593 

\bibitem[Gupta et al.(2005)]{gbd05} Gupta, S., B\"ottcher, M., \& Dermer, 
C.\ D.\ 2005, \apj, submitted

\bibitem[Hermsen et al.(1977)]{her77} Hermsen, W., et al.\ 
1977, \nat, 269, 494 

\bibitem[Jones(1968)]{j68} Jones, F.~C.\ 1968, Physical 
Review, 167, 1159 

\bibitem[Kaufman Bernad{\'o} et al.(2002)]{krm02} Kaufman 
Bernad{\'o}, M.~M., Romero, G.~E., \& Mirabel, I.~F.\ 2002, \aap, 385, L10 

\bibitem[Kniffen et al.(1997)]{kni97}Kniffen, D.~A., et al.\ 
1997, \apj, 486, 126  

\bibitem[Levinson \& Blandford(1996)]{lb96}Levinson, A.~\& 
Blandford, R.\ 1996, \apjl, 456, L29 

\bibitem[Li \& Wang(2004)]{lw04} Li, H., \& Wang, J.\ 2004, 
\apj, 617, 162 


\bibitem[Markoff et al.(2001)]{mff01} 
Markoff, S., Falcke, H., \& Fender, R.\ P.\ 2001,
A\&A, 372, L25

\bibitem[Markoff et al.(2003)]{mar03} Markoff, S., Nowak, M., 
Corbel, S., Fender, R., \& Falcke, H.\ 2003, \aap, 397, 645 

\bibitem[Massi(2004)]{mas04} 
Massi, M.\ 2004, \aap, 422, 267 

\bibitem[McSwain et al.(2001)]{mcs01} McSwain, M.~V., Gies, 
D.~R., Riddle, R.~L., Wang, Z., \& Wingert, D.~W.\ 2001, \apjl, 558, L43 

\bibitem[Melia \& K\"onigl(1989)]{mk89} Melia, F., \& K\"onigl, 
A.\ 1989, \apj, 340, 162 

\bibitem[Paredes(2005)]{par05} Paredes, J.~M.\ 2005, Chinese 
Journal of Astronony and Astrophysics, 5, 121 (astro-ph/0409226)

\bibitem[Paredes et al.(2000)]{par00} Paredes, J.~M., Mart{\'{\i}}, J., 
Rib{\' o}, M., \& Massi, M.\ 2000, Science, 288, 2340 

\bibitem[Paredes et al.(2002)]{par02} Paredes, J.~M., 
Rib{\'o}, M., Ros, E., Mart{\'{\i}}, J., \& Massi, M.\ 2002, \aap, 393, L99 

\bibitem[Paredes et al.(2005)]{pbr05}Paredes, J.\ M., Bosch-Ramon, V., 
\& Romero, G. E., 2005, A\&A, in press (astro-ph/0509095)

\bibitem[Protheroe et al.(1992)]{pmd92} Protheroe, R.~J., 
Mastichiadis, A., \& Dermer, C.~D.\ 1992, Astroparticle Physics, 1, 113 

\bibitem[Rachen \& M\'esz\'aros(1998)]{rm98} Rachen,  J.~P., and M{\' e}sz{\' a}ros, P.\ 1998, 
Phys.\ Rev.\ D, 58, 123005.

\bibitem[Rib{\'o} et al.(1999)]{rib99} Rib{\'o}, M., Reig, 
P., Mart{\'{\i}}, J., \& Paredes, J.~M.\ 1999, \aap, 347, 518 


\bibitem[Rieger(2004)]{rei04} Rieger, F.~M.\ 2004, \apjl, 
615, L5 

\bibitem[Romero et al.(2003)]{romero03}Romero, G. E., Torres, D. F., Kaufman Bernad\'o,
M. M., \& Mirabel, I. F., 2003, A\&A, 410, L1

\bibitem[Romero et al.(2005)]{rco05} Romero, G.~E., 
Christiansen, H.~R., \& Orellana, M.\ 2005, \apj, 632, 1093 

\bibitem[Romero(2005)]{rom05} Romero, G.~E.\ 2005, Chinese 
Journal of Astronony and Astrophysics, 5, 110 (astro-ph/0407461)

\bibitem[Sikora et al.(1994)]{sbr94} Sikora, M., Begelman, 
M.~C., \& Rees, M.~J.\ 1994, \apj, 421, 153 

\bibitem[Sikora(1997)]{sik97} Sikora, M.\ 1997, AIP 
Conf.~Proc.~410: Proceedings of the Fourth Compton Symposium, 410, 494 

\bibitem[Tavani et al.(1998)]{tav98} Tavani, M., Kniffen, D., 
Mattox, J.~R., Paredes, J.~M., \& Foster, R.\ 1998, \apjl, 497, L89 

\bibitem[Tavecchio et al.(1998)]{tave98} Tavecchio, F., 
Maraschi, L., \& Ghisellini, G.\ 1998, \apj, 509, 608 


\bibitem[Torres et al.(2003)]{tor03} Torres, D.\ F., Romero, G.\ E., 
Dame, T.\ M., Combi, J.\ A., and Butt, Y.\ M.\ 2003, Phys.\ Reports, 382, 303

\bibitem[Torres et al.(2005)]{tor05}Torres, D. F., Romero, G. E., Barcons, X.,
\& Lu, Y., 2005, ApJ, 626, 1015


\end{thebibliography}
\end{document}